\pdfoutput=1
\documentclass[useAMS,usenatbib]{mn2e}
\usepackage{amssymb}
\usepackage{bbm}
\usepackage{txfonts}
\usepackage{psfrag}
\usepackage{graphicx}
\usepackage{natbib}
\usepackage{if then}
\usepackage{color}
\usepackage{tikz}
\usetikzlibrary{shapes,arrows}
\usepackage{verbatim}

\def\del#1{{}}

\newcommand{\unit}[1]{\mbox{  }\rm{#1}}

\newcommand{\mat}[1]{{#1}}

\newcommand{\nextext}[1]{{{#1}}}

\title{Bayesian power-spectrum \nextext{inference} for Large Scale Structure data}
\author[Jens Jasche \\$^{1}$, Francisco Shu Kitaura, Benjamin D. Wandelt, Torsten A. En\ss lin]
       {Jens Jasche $^{1}$, Francisco S. Kitaura $^{2,1}$, Benjamin D. Wandelt $^{3}$ ,Torsten A. En\ss lin $^{1}$ \\$^{1}$ Max-Planck-Institut f\"{u}r Astrophysik , Karl-Schwarzschild Stra\ss e 1,  D-85748 Garching, Germany\\$^{2}$ SISSA, Scuola Internazionale Superiore di Studi Avanzati , via Beirut 2-4,  I-34014 Trieste, Italy\\$^{3}$ Department of Physics University of Illinois at Urbana-Champaign, 1110 West Green Street Urbana, IL 61801-3080, USA}
\begin{document}
\date{Submitted to MNRAS 22-May-2009}

\pagerange{\pageref{firstpage}--\pageref{lastpage}} \pubyear{2009}

\maketitle

\label{firstpage}

\begin{abstract}
We describe an exact, flexible, and computationally efficient algorithm for a joint estimation of the large-scale structure and its power-spectrum, building on a Gibbs sampling framework and present its implementation \textsc{ARES} (Algorithm for REconstruction and Sampling). \textsc{ARES} is designed to reconstruct the 3D power-spectrum together with the underlying dark matter density field in a Bayesian framework, under the reasonable assumption that the long wavelength Fourier components are Gaussian distributed. As a result \textsc{ARES} does not only provide a single estimate but samples from the joint posterior of the power-spectrum and density field conditional on a set of observations. This enables us to calculate any desired statistical summary, in particular we are able to provide joint uncertainty estimates. We apply our method to mock catalogs, with highly structured observational masks and selection functions, in order to demonstrate its ability to reconstruct the power-spectrum from real data sets, while fully accounting for any mask induced mode coupling. 
\end{abstract}


\section{Introduction}
\nextext{Throughout cosmic history a wealth of information on the origin and evolution of our Universe has been imprinted to the large scale structure via the gravitational amplification of primordial density perturbations. Harvesting this information from probes of the large scale structure, such as large galaxy surveys, therefore is an important scientific task to further our knowledge and to establish a conclusive cosmological picture. In recent years great advances have been made, both in retrieving huge amounts of data and increasing sensitivity in galaxy redshift surveys. Especially the recent galaxy surveys, the 2dF Galaxy Redshift Survey \citep[][]{COLLESS2001} and the Sloan Digital Sky Survey \citep[][]{SDSS7} provide sufficient redshifts to probe the 3D galaxy distribution on large scales.
In particular, the two point statistics of the matter distribution contains valuable information to test standard inflation and cosmological models, which describe the origin and evolution of all observed structure in the Universe.
Measuring the power-spectrum from galaxy observations therefore has always attracted great interest. Precise determination of the overall shape of the power-spectrum can for instance place important constraints on neutrino masses, help to identify the primordial power-spectrum, and break degeneracies for cosmological parameter estimation from CMB data \citep[e.g.][]{hu-98,wmap-spergel,HANNESTAD2003,Efstathiou_2002,PERCIVAL2002,wmap-spergel,VERDE2003}. In addition, several characteristic length scales have been imprinted to the matter distribution throughout cosmic history, which can serve as new standard rulers to measure the Universe.
A prominent example of these length scales is the sound horizon, which
yields oscillatory features in the power-spectrum, the so called
baryon \nextext{acoustic} oscillations (BAO) \citep[e.g.][]{SILK1968,PEEBLES1970,SUNYAEV1970}. Since the physics governing these oscillatory features is well understood, precise measurements of the BAO will allow us to establish a new, precise standard ruler to measure the Universe through the distance redshift relation \citep[][]{BLAKE2003,SEO2003}.
Precision analysis of large scale structure data therefore is a crucial step in developing a conclusive cosmological theory.}

\nextext{Unfortunately, contact between theory and observations cannot be made directly, since observational data is subject to a variety of systematic effects and statistical uncertainties. Such systematics and uncertainties arise either from the observational strategy or are due to intrinsic clustering behavior of the galaxy sample itself \citep[][]{SANCHEZ2008}. Some of the most prominent uncertainties and systematics are:}
\begin{itemize}
     \item survey geometry and selection effects
     \item close pair incompleteness due to fiber collisions
     \item galaxy biases
     \item redshift space distortions	
 \end{itemize}
\nextext{The details of galaxy clustering, and how galaxies trace the underlying density field are in general very complicated. The bias between galaxies and mass density is most likely non-linear and stochastical, so that the estimated galaxy spectrum is expected to differ from that of the mass \citep[][]{DEKEL1999}. Even in the limit where a linear bias could be employed, it still differs for different classes of galaxies \citep[see e.g.][]{COLE2005}. 
In addition, the apparent density field, obtained from redshift surveys, will generally be distorted along the line-of-sight due to the existence of peculiar velocities.}

\nextext{
However, the main cause for the systematic uncertainties in large scale power-spectrum estimations is the treatment of the survey geometry \citep[][]{TEGMARK1995,BALLINGER1995}. Due to the survey geometry the raw power-spectrum yields an expectation value for the power-spectrum, which is the true cosmic power-spectrum convolved with the survey mask \citep[][]{COLE2005}. This convolution causes an overall distortion of the power-spectrum shape, and drastically decreases the visibility of the baryonic features.}

\nextext{
The problems, mentioned above, have been discussed extensively in literature, and many different approaches to power-spectrum analysis have been proposed.
Some of the main techniques to recover the power-spectrum from galaxy surveys are Fourier transform based, such as the optimal weighting scheme, which assigns a weight to the galaxy fluctuation field, in order to reduce the error in the estimated power \citep[see e.g.][]{FELDMAN1994,TEGMARK1995,HAMILTON1997A,YAMAMOTO2003,PERCIVAL2004}. Alternative methods rely on Karhunen-Lo\`{e}ve decompositions \citep[][]{TEGMARK1997,TEGMARK_2004,POPE2004} or decompositions into spherical harmonics, which is especially suited to address the redshift space distortions problematic \citep[][]{FISHER1994,HEAVENS1995,TADROS1999,PERCIVAL2004,PERCIVAL2005}.
In addition, to these deconvolution methods there exists a variety of likelihood methods to estimate the real space power-spectrum \citep[][]{BALLINGER1995,HAMILTON1997A,HAMILTON1997B,TADROS1999,PERCIVAL2005}. In order to not just provide the maximum likelihood estimate but also conditional errors, \cite{PERCIVAL2005} proposed a Markov Chain Monte Carlo method to map out the likelihood surface. 
}

\nextext{
Nevertheless, as the precision of large scale structure experiments has improved, the requirement on the control and characterization of systematic effects, as discussed above, also steadily increases. It is of critical importance to propagate properly the uncertainties caused by these effects through to the matter power-spectrum and cosmological parameters estimates, in order to not underestimate the final uncertainties and thereby draw incorrect conclusions on the cosmological model.
}

\nextext{We therefore felt inspired to develope a new Bayesian approach to extract information on the two point statistics from a given large scale structure dataset. We prefer Bayesian methods to conventional likelihood methods, as the yield more general and profound statements about measurements. In example, conventional likelihood methods can only answer questions of the inner form like :" Given the true value \(s\) of a signal, what is the probability distribution of the measured values \(d\)?" A Bayesian method, on the other hand, answers questions of the type :"Given the observations \(d\), what is the probability distribution of the true underlying signal \(s\)?" For this reason, Bayesian statistics answers the underlying question to every measurement problem, of how to estimate the true value of the signal from observations, while conventional likelihood methods do not \citep[][]{MICHEL1999}.}
\nextext{Since the result of any Bayesian method is a complete probability distribution they permit fully global analyses, taking into account all systematic effects and statistical uncertainties. In particular, here, we aim at evaluating the power-spectrum posterior distribution \(\mathcal{P}\left(\{P(k_i)\}|\{d_i\}\right)\), with \(P(k_i)\) being the power-spectrum coefficients of the \(k_i\)th mode and \(d_i=d(\vec{x}_i)\) is an observation at position \(\vec{x}_i\) in three dimensional configuration space.}
 This probability distribution would then contain all information on the two point statistics supported by the data. 
In order to explore this posterior distribution we employ a Gibbs sampling method, previously applied to CMB data analysis \citep[see e.g.][]{WANDELT2004,2004ApJS..155..227E,JEWELL2004}.

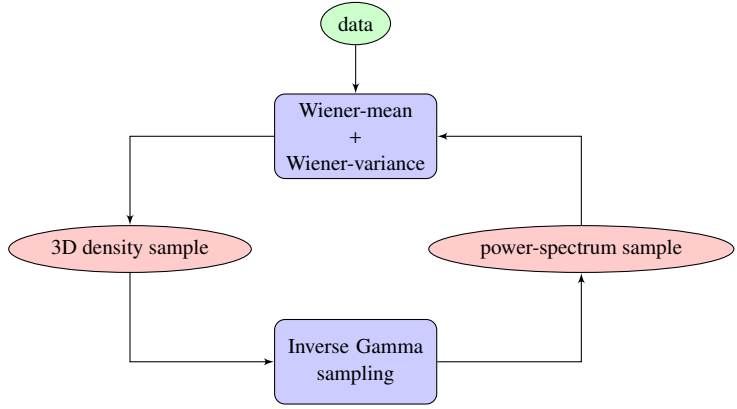
\begin{figure*}
	\centering
	{
	\tikzstyle{blank} = [rectangle,  fill=white!20,text width=5em, text centered, rounded corners, minimum height=4em]
	\tikzstyle{decision} = [diamond, draw, fill=blue!20,text width=4.5em, text badly centered, node distance=3cm, inner sep=0pt]
	\tikzstyle{block} = [rectangle, draw, fill=blue!20,text width=7em, text centered, rounded corners, minimum height=4em]
	\tikzstyle{line} = [draw, -latex']
	\tikzstyle{cloud} = [draw, ellipse,fill=red!20, node distance=3cm, minimum height=2em]
	\tikzstyle{clouda} = [draw, ellipse,fill=green!20, node distance=3cm, minimum height=2em]
\begin{tikzpicture}[node distance = 1.5cm, auto]
    % Place nodes
    \node [clouda] (data) {data};
   \node [block,below of=data] (DENSSAMPLING) {Wiener-mean \\ + \\ Wiener-variance};
    \node [blank,below of=DENSSAMPLING] (blank) {};
    \node [block,below of=blank] (PSSAMPLING) {Inverse Gamma sampling};	
    \node [cloud,left of=blank] (denssample) {3D density sample};
    \node [cloud,right of=blank] (pssample) {power-spectrum sample};	
    \path [line] (data) -- (DENSSAMPLING);
    \path [line] (DENSSAMPLING) -| (denssample);
    \path [line] (denssample) |- (PSSAMPLING);	
    \path [line] (PSSAMPLING) -| (pssample);
    \path [line] (pssample) |- (DENSSAMPLING);
\end{tikzpicture}
	}
\caption{Flow-chart depicting the two step iterative Gibbs sampling procedure.}
	\label{fig:flowchart}
\end{figure*}

\nextext{
Since direct sampling from \(\mathcal{P}\left(\{P(k_i)\}|\{d_i\}\right)\) is impossible or at least difficult, they propose instead to draw samples from the full joint posterior distribution \(\mathcal{P}\left(\{P(k_i)\},\{s_i\}|\{d_i\}\right)\) of the power-spectrum coefficients \(P(k_i)\) and the 3D matter density contrast amplitudes \(s_i\) conditional on a given set of data points \(\{d_i\}\).
This is achieved by iteratively drawing density samples from a Wiener-posterior distribution and power-spectrum samples via an efficient Gibbs sampling scheme (see figure \ref{fig:flowchart} for an illustration). Here, artificial mode coupling, as introduced by survey geometry and selection function, is resolved by solving the Wiener-filtering equation, which naturally regularizes inversions of the observational response operator in unobserved regions.}
In this fashion\nextext{,} we obtain a set of Monte Carlo samples from the joint posterior, which allows us to compute any desired property of the joint posterior density, with the accuracy only limited by the sample size.
In particular\nextext{,} we obtain \nextext{the power spectrum posterior} \(\mathcal{P}\left(\{P(k_i)\}|\{d_i\}\right)\) by simply marginalizing \nextext{the joint posterior} \(\mathcal{P}\left(\{P(k_i)\},\{s_i\}|\{d_i\}\right)\) over the auxiliary density amplitudes \(s_i\), which is trivially achieved by ignoring the \(s_i\) samples.

The Gibbs sampler also offers unique capabilities for propagating systematic uncertainties end-to-end. Any effect, for which we can define a sampling algorithm, either jointly with or conditionally on other quantities, can be propagated seamlessly through to the final posterior. 

It is worth noting, that our method differs from traditional methods of analyzing galaxy surveys in a fundamental aspect. 
Traditional methods consider the analysis task as a set of steps, each of which arrives at intermediate outputs which are then fed as inputs to the next step in the pipeline. Our approach is a truly global analysis, in the sense that the statistics of all science products are computed jointly, respecting and exploiting the full statistical dependence structure between various components.

In this paper we present \textsc{ARES} (Algorithm for REconstruction and Sampling), a computer algorithm to perform a full Bayesian data analysis of 3D redshift surveys. 
In section \ref{LSS_SAMPLER} we give an introduction to the general idea of the large scale structure Gibbs sampling approach, followed by section \ref{MAP_SAMPLER} and \ref{PS_SAMPLER}, where we describe and derive in detail the necessary ingredients to sample the 3D density distribution and the power-spectrum respectively. The choice of the prior and the relevance for the cosmic variance are discussed in section \ref{Prior_and_Variance}.
Details concerning the numerical implementation are discussed in section \ref{NUMERICAL_IMPLEMENTATION}. We then test \textsc{ARES} thouroughly in section \ref{Gaussian_Test_Cases}, particularly focussing on the treatment of survey masks and selection functions. In section \ref{OPERATIONS_GIBBS_SAMPLES} we demonstrate the running median filter, and use it as an example to demonstrate how uncertainties can be propagated to all inferences based on the set of Gibbs samples. Finally we conclude in section \ref{Conclusion}, by discussing the results of the method and giving an outlook for future extensions and application of our method.

\section{Notation}
\label{Notation}
In this section, we describe the basic notation used throughout this work. Let the quantity \(\rho_i=\rho(\vec{x}_i)\) be the field amplitude of the three dimensional field \(\rho(\vec{x})\) at position \(\vec{x}_i\). Then the index \(i\) has to be understood as a multi index, which labels the three components of the position vector:
\begin{equation}
\label{eq:multi_index}
\vec{x}_i =[x^1_i,x^2_i,x^3_i] \, ,
\end{equation}
where \(x^j_i\) is the \(j\)th component of the \(i\)th position vector. Alternatively one can understand the index \(i\) as a set of three indices \(\{r,s,t\}\) so that for an equidistant grid along the three axes the position vector can be expressed as:
\begin{equation}
\label{eq:multi_index_a}
\vec{x}_i =\vec{x}_{r,s,t} = [\Delta x\, r,\Delta y\, s,\Delta z\, t] \, ,
\end{equation}
with \(\Delta x\), \(\Delta y\) and \(\Delta z\) being the grid spacing along the three axes.
With this definition we yield:
\begin{equation}
\label{eq:multi_index_a}
\rho_i \equiv \rho_{r,s,t} \, .
\end{equation}
Also note that any summation running over the multi index \(i\) is defined as the three sums over the three indices \(r\), \(s\) and \(t\):
\begin{equation}
\label{eq:multi_index_c}
\sum_i \equiv \sum_r \sum_s \sum_t \, .
\end{equation}
Further, we will frequently use the notation \(\{\rho_i\}\), which denotes the set of field amplitudes at different positions \(\vec{x}_i\).
In particular:
\begin{equation}
\label{eq:multi_index_set}
\{\rho_i\} \equiv \{\rho_0, \rho_1, \rho_2, ... ,\rho_{N-1}\} \, ,
\end{equation}
where \(N\) is the total number of position vectors.

\section{The Large scale structure Gibbs sampler}
\label{LSS_SAMPLER}
As already described in the introduction, we seek to sample from the joint posterior distribution \(\mathcal{P}\left(\{P(k_i)\},\{s_i\}|\{d_i\}\right)\) of the power-spectrum coefficients \(P(k_i)\) and the 3D matter density contrast amplitudes \(s_i\) given a set of observations \(\{d_i\}\).

In principle, this joint posterior distribution could be mapped out over a grid in the multi-dimensional space of the signal amplitudes \(s_i\) and power-spectrum coefficients \(P(k_i)\).
But since the number of grid points required for such an analysis scales exponentially with the number of free parameters, this approach cannot be realized efficiently.
For this reason, we propose a Gibbs sampling approach to this problem.

The theory of Gibbs sampling \citep{Gelfand_1990,Tanner_1996,OHagan} states, that if it is possible to sample from the conditional densities \({\cal P}(\{s_i\}|\{P(k_i)\},\{d_i\})\) and \({\cal P}(\{P(k_i)\}|\{s_i\},\{d_i\})\), then iterating the following two sampling equations will, after an initial burn-in period, lead to samples from the joint posterior \({\cal P}(\{s_i\},\{P(k_i)\}|\{d_i\})\):
\begin{equation}
\label{eq:signal_sampling}
\{s_i\}^{(j+1)}\curvearrowleft {\cal P}(\{s_i\}|\{P(k_i)\}^{(j)},\{d_i\}) \, ,
\end{equation}
\begin{equation}
\label{eq:Pspec_sampling}
\{P(k_i)\}^{(j+1)}\curvearrowleft {\cal P}(\{P(k_i)\}|\{s_i\}^{(j+1)},\{d_i\}) \, ,
\end{equation}
where the symbol \(\curvearrowleft\) denotes a random draw from the probability density on its right.

Once a set of samples from \({\cal P}(\{s_i\},\{P(k_i)\}|\{d_i\})\) has been obtained, the properties of this probability density can be summarized in terms of any preferred statistic, such as its multivariate mean, mode or variance.

As our approach probes the joint distribution, we are able to quantify joint uncertainties of the signal amplitudes and the power-spectrum conditional just on the data. For this reason, the Gibbs sampling approach should not be considered as yet another maximum likelihood technique, although it certainly is able to produce such an estimate.

In the following we are going to describe the necessary methods and procedures required for iterating the processes \ref{eq:signal_sampling} and \ref{eq:Pspec_sampling} of signal and power-spectrum sampling.

\section{Sampling the signal maps}
\label{MAP_SAMPLER}
Assuming a Gaussian signal posterior distribution \({\cal P}(\{s_i\}|\{P(k_i)\},\{d_i\})\), the task of drawing a random signal sample can be split into two steps.

First, we estimate the maximum a postiori values for the signal amplitudes \(s_i\), which in the Gaussian case coincide with the mean values.
Then a fluctuation term, being consistent with the correct covariance, is added to the mean signal. The sum of the mean and the fluctuation term will then yield a sample from the conditional posterior.

The most challenging procedure in this signal sampling step is to calculate the a postiori values for the signal amplitudes \(s_i\). 
Assuming a Gaussian posterior will directly lead to a Wiener filtering procedure, described below.
However, this method requires to invert huge matrices which consists of the sum of the inverse signal \(\mat{S}\) and inverse noise \(\mat{N}\) covariance matrices. The matrix inversion is a numerically very demanding step, and at the same time presents the bottleneck for our method, as it defines the computational speed with which a signal sample can be produced.
The efficient implementation of these matrix inversion step, as described by \cite{Kitaura}, allows for the production of many thousands of samples in computational feasible times.
In the following sections we will describe the details of the signal sampling procedure.

\subsection{The Wiener filter}
\label{WIENER_FILTER}
As already described above, the main task for the signal sampling step is to derive the maximum a postiori values for the signal amplitudes \(s_i\).
According to Bayes' theorem the conditional signal posterior can be written as the product of a signal prior and a likelihood normalized by the so called evidence. Further, here we will use the signal covariance matrix \(\mat{S}\) rather than the power-spectrum \(\{P(k_i)\}\).  It is well known, that the power-spectrum is just the Fourier transform of the signal covariance in configuration space. Since the Fourier transform is a basis transformation with a unitary transformation matrix, the signal covariance matrix \(\mat{S}\) and the power-spectrum \(\{P(k_i)\}\) can be used interchangeably for a normalized Fourier transform (see section \ref{Power_spectrum_sampling} and Appendix \ref{CHANGE_TO_FFT_REPRESENTATION} for more details).

We can therefore write the signal posterior as:
\begin{eqnarray}
\label{eq:signal_posterior}
{\cal P}(\{s_i\}|\mat{S},\{d_i\})&=&\frac{{\cal P}(\mat{S})}{{\cal P}(\{d_i\},\mat{S})}\,{\cal P}(\{s_i\}|\mat{S})\,{\cal P}(\{d_i\}|\{s_i\},\mat{S}) \nonumber\\
&=& \frac{1}{{\cal P}(\{d_i\}|\mat{S})}\,{\cal P}(\{s_i\}|\mat{S})\,{\cal P}(\{d_i\}|\{s_i\}) \, ,
\end{eqnarray}
where we assume that the data amplitudes \(d_i\) are conditionally independent of the signal covariance matrix \(\mat{S}\), once the signal amplitudes \(s_i\) are given.
Following \cite{BARDEEN1986}, we describe the signal prior for the large scale matter distribution as a multivariate Gaussian, with zero mean and the signal covariance \(\mat{S}\). We can then write:
\begin{equation}
\label{eq:signal_prior}
{\cal P}(\{s_i\}|\mat{S})=\frac{1}{\sqrt{\rm{det}\left(2\pi \mat{S}\right)}} e^{-\frac{1}{2}\sum_i\sum_j \,s_i\, \mat{S_{ij}}^{-1}\,s_j} \, .
\end{equation}
The Fourier transform of the signal covariance matrix \(\mat{S}\) has an especially appealing form in Fourier space.
It is well known, that in an homogeneous and isotropic universe the Fourier transform of the signal covariance is a diagonal matrix, with the diagonal elements being the power-spectrum. Hence, we can express the Fourier representation of the signal covariance as:
\begin{equation}
\label{signal_covarianceFS}
\hat{\hat{\mat{S}}}_{kl}= \delta^K_{kl}\, P_k \,
\end{equation}
where the \(\,\hat{}\,\)-symbol denotes a Fourier transform, \(\delta^K_{ij}\) is the Kronecker delta and \(P_k=P(k_k)\) is the power-spectrum coefficient at the Fourier mode \(\vec{k}_k\) in three dimensional pixel space \citep[see e.g.][]{PADMANABHAN1993,2004LRR.....7....8L}. 

The choice of the Gaussian prior can be justified by inflationary theories, which predict the matter field amplitudes to be Gaussian distributed in the linear regimes at scales \(k \lesssim 0.15 \, \rm{h/Mpc}\) \citep[][]{PEACOCK1994,PERCIVAL2001}. 

%As already described in the introduction, inflationary models predict the initial density field to obey Gaussian statistics. In this case the above prior would completely describe the full statistics of the matter field in the universe.

%However, these initial seed fluctuations have undergone gravitational collapse to form the structures observed in the sky today. This process in general does not leave the initial statistics of the density field intact, so that the statistics of the density field evolved away from a Gaussian distribution.

%However, it is also known, that the largest scales still follow the linear gravitational evolution, which allows us to relate the present density field at the largest scales with the initial density field by simply scaling it with a proportionality constant, the so called growth factor.

%It is therefore believed, that at the linear scales the Gaussian prior is still the adequate representation of the statistical behavior of the matter distribution \citep[][]{PEACOCK1994}.

At nonlinear scales the Gaussian prior does not represent the full statistical behavior of the matter field anymore. During nonlinear gravitational structure formation the statistics of the initial density field has evolved from a Gaussian distribution towards a log normal distribution as commonly assumed in literature \citep[][]{COLES1991,COLOMBI1994,KAYO2001}.

However, note that in this case the Gaussian prior still describes the two point statistics of the underlying density field even in the nonlinear regime. The Gaussian prior should therefore be regarded as our a priori knowledge of the matter distribution, which is only formulated up to two point statistics. Next, we discuss the likelihood  \({\cal P}(\{d_i\}|\{s_i\})\) given in equation (\ref{eq:signal_posterior}).

As we seek to recover the maximum a postiori signal \(s_i\) from the set of observations \(d_i\) we must assume a model which relates these both quantities. The most straight forward data model is linear, and can be written as:
\begin{equation}
\label{eq:WIENER_DATA_MODEL}
d_i=\sum_k \, K_{ik}\, s_k + \epsilon_i \, ,
\end{equation}
where \(K_{ij}\) is an observation response operator and \(\epsilon_i\) is an additive noise contribution, which will be defined in more detail in the next section. 
If we assume the noise \(\epsilon_i\) to be Gaussian distributed, with zero mean and covariance \(\mat{N}\), we can express the likelihood as:
\begin{equation}
\label{signal_likelihood}
{\cal P}(\{d_i\}|\{s_i\})=\frac{1}{\sqrt{\rm{det}\left(2\pi \mat{N}\right)}} e^{-\frac{1}{2}\left(\sum_i\sum_j \,\left[d_i-\sum_k \, K_{ik}\, s_k\right]\, \mat{N_{ij}}^{-1}\,\left[d_j-\sum_l \, K_{jl}\, s_l\right]\right)} \, ,
\end{equation}
where we simply inserted the data model given in equation (\ref{eq:WIENER_DATA_MODEL}) into the Gaussian noise distribution.

With these definitions the signal posterior is a multivariate Gaussian distribution and can be written as:
\begin{eqnarray}
\label{eq:Gaussian_signal_posterior}
{\cal P}(\{s_i\}|\mat{S},\{d_i\})&\propto& e^{-\frac{1}{2}\left(\sum_i\sum_j \,s_i\, \mat{S_{ij}}^{-1}\,s_j + \,\left[d_i-\sum_k \, K_{ik}\, s_k\right]\, \mat{N_{ij}}^{-1}\,\left[d_j-\sum_l \, K_{jl}\, s_l\right]\right)}\, , \nonumber \\
\end{eqnarray}
where we omitted the normalization factor, which is of no interest in the following.

Note, that omitting the normalization of the likelihood, requires that the additive noise term is independent of any signal contribution, as otherwise the noise covariance matrix would carry signal information and could not be neglected in the following. This assumption, however, is in general not true for the Poissonian noise, as described, in the next section.

The maximum of this signal posterior can then easily be found by either completing the square in the exponent of equation (\ref{eq:Gaussian_signal_posterior}), or by extremizing \({\cal P}(\{s_i\}|\mat{S},\{d_i\})\) with respect to the signal amplitudes \(s_i\).
The latter approach allows us to directly read of the Wiener filter equation from equation (\ref{eq:Gaussian_signal_posterior}), by simply differentiating the exponent with respect to \(s_i\) and setting the result to zero.
The result is the famous Wiener filter equation which is given as: 
\begin{eqnarray}
\label{eq:WIENER_FILTER_EQUATION}
\sum_j \left[ \mat{S}^{-1}_{ij} + \sum_m \sum_l \mat{K}_{mi} \mat{N}^{-1}_{ml} \mat{K}_{lj} \right] \, m_j = \sum_m \sum_l \mat{K}_{mi} \mat{N}^{-1}_{ml} d_l \, ,
\end{eqnarray}
where we denoted the variable \(m_j\) as a Wiener mean signal amplitude, to clarify that this reconstruction is the mean and not a typical sample of the distribution.
The solution of this equation requires to invert the matrix:
\begin{equation}
\label{eq:LSS_Posterior_1}
\mat{D}_{ij} = \mat{S}^{-1}_{ij} + \sum_m \sum_l \mat{K}_{mi} \mat{N}^{-1}_{ml} \mat{K}_{lj} \, ,
\end{equation}
which leads to the solution for the signal amplitudes
\begin{eqnarray}
\label{eq:WIENER_SOLUTION}
m_i = \sum_j \mat{D}^{-1}_{ij}\sum_m \sum_l \mat{K}_{mj} \mat{N}^{-1}_{ml} d_l \, .
\end{eqnarray}
This result demonstrates that estimating the maximum a postiori values \(m_i\) for the signal amplitudes \(s_i\) involves inversions of the Wiener filter operator \(\mat{D}\).
Therefore, the signal-sampling operation is by far the most demanding step of our Gibbs sampling scheme, as it requires the solution of a very large linear system.

Formally speaking, in practice, this corresponds to inverting matrices of order \(\sim 10^6 \times 10^6\) or larger, which clearly is not computationally feasible through brute-force methods. For example, matrix inversion algorithms, based on usual linear algebra methods, have a numerically prohibitive \(\mathcal{O}(N_{pix}^3)\) scaling, in order to transform to the eigenspace of the system, which bars sampling from the signal posterior.

This is the situation in which \cite{Kitaura} proposed a particular operator based inversion technique to allow for computationally efficient calculation of the Wiener filter equation in three dimensional space.

In this implementation, the system of equations (\ref{eq:WIENER_SOLUTION}) can be solved by means of conjugate gradients (CGs). The computational scaling of this method is thus reduced to the most expensive step for applying the operator on the right-hand side of the equations, which in our case is the Fast Fourier transform, which scales as \(\mathcal{O}(N_{pix}\,\log(N_{pix}))\) .

\subsection{The galaxy data model}
\label{DATA_MODEL}
In order to adapt the Wiener filter procedure for the specific application to galaxy observations, we are going to present the galaxy data model together with the according Poissonian noise covariance matrix.

It is possible to describe the observed galaxy distribution as a realization of an inhomogeneous Poissonian process \citep[][]{MARTINEZ2002}.
We can therefore assume the observed galaxy numbers \(N^{O}_i=N^{O}(\vec{x}_i)\) at position \(\vec{x}_i\) in three dimensional configuration space to be drawn from a Poissonian distribution \citep[][]{MARTINEZ2002,KITAURA2009}.
\begin{equation}
\label{eq:Poissonian}
N^{O}_i\curvearrowleft {\mathcal P}(N^{O}_i|\lambda^{O}_i)= \frac{{\lambda^{O}_i}^{N^{O}_i} e^{-\lambda^{O}_i}}{{N^{O}_i}!} \, ,
\end{equation}
where the arrow denotes a random draw from the probability distribution and \(\lambda^{O}_i\) is the mean observable galaxy number at position \(\vec{x}_i\).
We can then write the observed galaxy numbers at discrete positions as:
\begin{equation}
\label{eq:Observed_gal_num}
N^{O}_i= \langle N^{O}_i \rangle + \epsilon^{O}_i = \lambda^{O}_i + \epsilon^{O}_i  \, ,
\end{equation}
where the noise term \(\epsilon^{O}_i\) denotes the difference between the observed galaxy number and the mean observable galaxy number.
The Poissonian noise covariance matrix can then easily be obtained by:
\begin{equation}
\label{eq:Poissonian_Noise_Cov}
N^{P}_{ij} = \langle \epsilon^{O}_i \epsilon^{O}_j \rangle = \langle [N^{O}_i -\langle N^{O}_i \rangle][N^{O}_j -\langle N^{O}_j \rangle] \rangle=\delta^K_{ij}\langle N^{O}_i \rangle =\delta^K_{ij}\,\lambda^{O}_i \, ,
\end{equation}
where we simply calculated the Poissionian variance for the observed galaxy number assuming the galaxies to be independent and identically distributed (i.i.d).
The mean observable galaxy number can be related to the true mean galaxy number \(\lambda_i\) by applying the observation response operator \(R_{ij}\) as:
\begin{equation}
\label{eq:True_gal_num}
\lambda^{O}_{i} = \sum_j\, R_{ij}\,\lambda_j\, ,
\end{equation}
The true mean galaxy number, on the other hand, can be related to the dark matter over density field, the signal \(s_i\), by introducing a physical model in the form of a bias operator \(B_{ij}\), e.g. a scale dependent bias:
\begin{equation}
\label{eq:True_signal}
\lambda_i = \bar{\lambda} \left(1+\sum_j B_{ij}\, s_j \right) \, .
\end{equation}
By inserting equations (\ref{eq:True_gal_num}) and (\ref{eq:True_signal}) into equation (\ref{eq:Observed_gal_num}) and applying trivial algebraic conversions, we yield the data model: 
\begin{eqnarray}
\label{eq:data_model}
d_i &=& \frac{N^{O}_i}{\bar{\lambda}}-\sum_j\, R_{ij}= \sum_j\, R_{ij} \sum_k \, B_{jk}\, s_k + \frac{\epsilon^{O}_i}{\bar{\lambda}} \, ,
\end{eqnarray}
For the case of galaxy redshift surveys the response operator \(R_{ij}\) is the product of the sky mask and the selection function, which are both local in configuration space, and hence the response operator turns to:
\begin{equation}
\label{eq:Response_operator}
R_{ij} = \delta^K_{ij} M_i\,F_i  \, ,
\end{equation}
where \(M_i\) is the value of the sky mask and \(F_i\) is the value of the selection function at position \(i\).
We therefore arrive at the data model already described in equation (\ref{eq:WIENER_DATA_MODEL}), which reads:
\begin{eqnarray}
\label{eq:data_model_a}
d_i &=& M_i\,F_i \sum_k \, B_{ik}\, s_k + \frac{\epsilon^{O}_i}{\bar{\lambda}} \nonumber \\
&=& \sum_k \, K_{ik}\, s_k + \epsilon_i \, ,
\end{eqnarray}
where we introduced the effective observation response operator \(K_{ij}=M_i\,F_i\, B_{ij}\) and the noise contribution \( \epsilon_i= \epsilon^{O}_i/\bar{\lambda}\).
This is the galaxy data model which we derived from the assumption of the Poissonian distribution of galaxies.

The Wiener filter operator requires the definition of the noise covariance matrix \(\mat{N}\), which for the Poissonian noise can be expressed as:
\begin{equation}
\label{Noise_Cov}
N_{ij} = \langle \epsilon_i \epsilon_j \rangle= \frac{\langle \epsilon^{O}_i \epsilon^{O}_j \rangle}{\bar{\lambda}^2}=\delta^K_{ij}\,\frac{\lambda^{O}_i}{\bar{\lambda}^2} \, ,
\end{equation}
where we used the Poissonian noise covariance matrix given in equation (\ref{eq:Poissonian_Noise_Cov}).

However, introducing equation (\ref{eq:True_gal_num}) and (\ref{eq:True_signal}) yields the noise covariance matrix:
\begin{equation}
\label{Noise_Cov}
\mat{N}_{ij} =\delta^K_{ij}\,\frac{1}{\bar{\lambda}} \,\left [\sum_k\mat{R}_{ik} \left(1+\sum_{l}\mat{B}_{kl}\,s_l\right)\right]\, ,
\end{equation}
which immediately reveals, that there is a correlation between the underlying signal amplitudes \(s_i\) and the level of shot noise produced by the discrete distribution of galaxies \citep[see e.g.][]{1998ApJ...503..492S}.

Nevertheless, as pointed out in the previous section, the Wiener filter relies on the fact, that the additive noise contribution is uncorrelated with the signal.
Hence, we have to assume the noise covariance as uncorrelated with the signal, but it may have some structure.

Therefore, we provide two approaches to effectively approximate the noise covariance matrix given in equation (\ref{Noise_Cov}).

\begin{figure*}
\centering{\resizebox{1.\hsize}{!}{\includegraphics{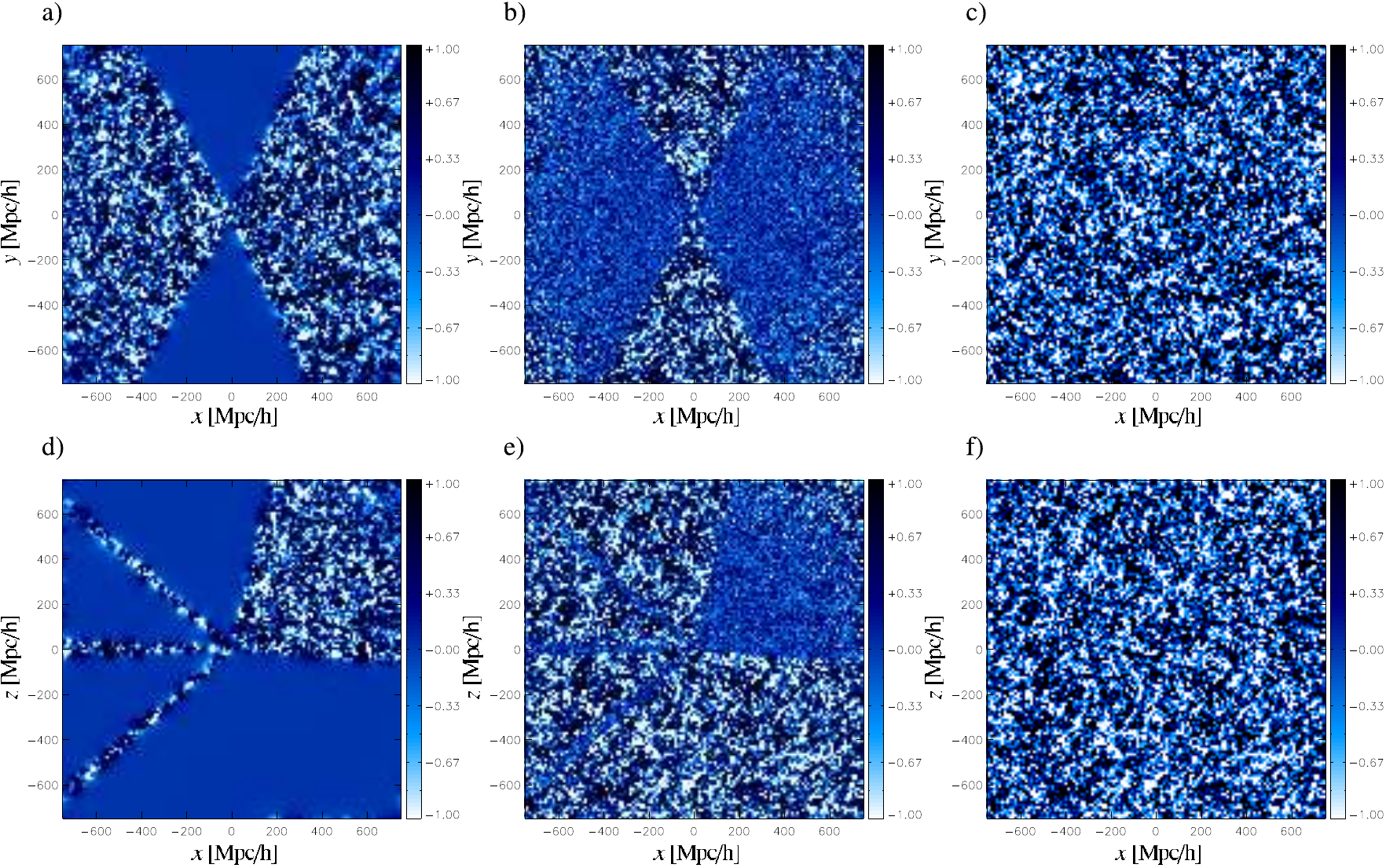}}}
\caption{Slices through signal sample produced in one Gibbs sampling step. The left panels (a,d) show the Wiener filtered mean signal, panels (b,e) present the fluctuation term, and the right panels show the full, noiseless Gibbs sample. The color map encodes the amplitude of the density contrast.}
\label{fig:WIENER_VARIANCE}
\end{figure*}

In the first approach we calculate an effective noise covariance matrix by averaging the noise covariance matrix given in equation (\ref{Noise_Cov}) over the signal.
We then obtain:
\begin{eqnarray}
\label{Noise_Cov_av}
\bar{\mat{N}}_{ij} &= & \langle \mat{N}_{ij} \rangle _{s} \nonumber \\
&=& \delta^K_{ij}\,\frac{1}{\bar{\lambda}} \,\left [\sum_k\mat{R}_{ik} \left(1+\sum_{l}\mat{B}_{kl}\,\langle s_l\rangle _{s} \right)\right]\nonumber \\
&=& \delta^K_{ij}\,\frac{1}{\bar{\lambda}} \,\left [\sum_k\mat{R}_{ik}\right]\, ,
\end{eqnarray}
where we used the fact, that the ensemble mean of the signal amplitudes for the density contrast vanishes. Note, that this model also arises when persuing a least squares approach to matter field reconstructions rather than the Bayesian approach as described in this work \citep[for details see][]{Kitaura}.

In the other approach we introduce a noise structure function \(n^{SF}_{i}\) given as:
\begin{equation}
\label{eq:noisestructurefunction}
n^{SF}_{i}=\frac{\lambda^{O}_i}{\bar{\lambda}^2} \, .
\end{equation}
With this definition the noise is approximated as being uncorrelated to the signal, but nonuniform. The noise covariance matrix then reads:
\begin{equation}
\label{Noise_Cov_SF}
\mat{N}^{SF}_{ij} = \delta^K_{ij}\,n^{SF}_{i} \, .
\end{equation}
In order to use this noise structure function we have to estimate \(\lambda^{O}_i\) from the observed galaxy numbers \(N^{O}_i\).
By applying Bayes' theorem to the Poissonian distribution given in relation (\ref{eq:Poissonian}) we yield:
\begin{equation}
\label{Noise_Cov_SF_PDF}
{\mathcal P}(\lambda^{O}_i|N^{O}_i) = {\mathcal P}(\lambda^{O}_i)\,\frac{{\mathcal P}(N^{O}_i|\lambda^{O}_i)}{{\mathcal P}(N^{O}_i)} \, .
\end{equation}
In the absence of any further a priori knowledge on \(\lambda^{O}_i\) we assume a flat prior and search for the maximum of:
\begin{equation}
\label{eq:Noise_Cov_SF_PDF_a}
{\mathcal P}(\lambda^{O}_i|N^{O}_i) = \frac{{\lambda^{O}_i}^{N^{O}_i} e^{-\lambda^{O}_i}}{\Gamma (N^{O}_i+1)} \, ,
\end{equation}
which is normalized to yield unity when integrated over all \({\lambda^{O}_i}\).

The noise structure function \(n^{SF}_{i}\) can then be estimated by searching the most likely value for \(\lambda^{O}_i\) from equation (\ref{eq:Noise_Cov_SF_PDF_a}). This yields: 
\begin{equation}
\label{NOISE_SF_MAX}
n^{SF}_{i} = \frac{N^{O}_i}{\bar{\lambda}^2}  \, .
\end{equation}
Another estimator for \(n^{SF}_{i}\) is based on evaluating the mean of the probability distribution given in equation (\ref{eq:Noise_Cov_SF_PDF_a}).
The ensemble mean is calculated as:
\begin{equation}
\label{NOISE_SF_MEAN_a}
\langle \lambda^{O}_i \rangle = \int_0^{\infty} d\lambda^{O}_i \lambda^{O}_i \frac{{\lambda^{O}_i}^{N^{O}_i} e^{-\lambda^{O}_i}}{\Gamma (N^{O}_i+1)}=N^{O}_i+1  \, .
\end{equation}
in this case the noise structure function \(n^{SF}_{i}\) can be written as:
\begin{equation}
\label{NOISE_SF_MEAN_b}
n^{SF}_{i} = \frac{N^{O}_i+1}{\bar{\lambda}^2}  \, .
\end{equation}
A more thorough discussion on Poissonian noise models and their numerical implications for matter field reconstructions can be found in \cite{KITAURA2009}.

\subsection{Drawing signal samples}
\label{DRAWING_SIGNAL_SAMPLES}
In the previous sections, we described the Wiener filter and the galaxy data model, which are required to estimate the mean of the signal posterior.
However, this mean signal is no sample from the signal posterior yet, neither does it represent a physical density field, as it lacks power in the low signal to noise regions.
To create a true sample from the signal posterior, one must therefore add a fluctuation term \(y_i\), which compensates the power lost due to noise filtering.
The signal sample can then be written as the sum of the signal mean and a fluctuation term:
\begin{equation}
\label{eq:AUGMENT_EQUATION}
s_i = m_i + y_i \, .
\end{equation}
In our approach we realize the fluctuation term by generating a mock signal \(s_i^*\) and a mock observation \(d_i^*\) consistent with the data model given in equation (\ref{eq:data_model_a}). This kind of mock observation generation is well known in literature and has been applied to various scientific applications, as for instance the generation of constrained initial conditions for Nbody simulations \citep[see e.g.][]{Bertschinger_1987,Hoffman_1991ApJ,Ganon_1993,Kitaura}. The fluctuation term can then simply be calculated as:
\begin{equation}
\label{eq:FLUCTUATION_TERM}
y_i= s_i^*-\sum_j \mat{D}^{-1}_{ij}\sum_m \sum_l \mat{K}_{mj} \mat{N}^{-1}_{ml} d_l^* \, .
\end{equation}
The interpretation of this equation is simple. In the high signal to noise regime, the Wiener filter is nearly a pass-through operator, meaning the reconstructed signal is nearly identical to the true underlying signal. Therefore, as the variance is low, the fluctuation term tends towards zero.
In the low signal to noise regime, on the other hand, the Wiener filter will block, and no signal can be reconstructed. The fluctuation term will therefore be nearly identical to the underlying mock signal \(s_i^*\). 

In this fashion we add the correct power to the Wiener mean reconstruction. The effect of adding the fluctuation term to the Wiener mean is presented in figure \ref{fig:WIENER_VARIANCE}, where we see the Wiener mean reconstruction, the fluctuation term and the sum of both.

The mock data \(d_i^*\) is generated to obey the data model described in equation (\ref{eq:data_model_a}) and the Wiener variance.

We therefore first draw a mock signal \(s_i^*\) with correct statistics from the multivariate Gaussian signal prior given in equation (\ref{eq:signal_prior}). Such a mock signal is best generated in Fourier space following the description of \cite{2005astro.ph..6540M}. One first draws two Gaussian random numbers, \(\chi_a\) and \(\chi_b\), with zero mean and unit variance and then calculates the real and imaginary part of the signal in Fourier space as:
\begin{eqnarray}
\label{eq:SIGNAL_GENERATION}
RE(\hat{s}_k) &=& \sqrt{\frac{P_k}{2}}\, \chi_a \nonumber \\
IM(\hat{s}_k) &=& \sqrt{\frac{P_k}{2}}\, \chi_b \, ,
\end{eqnarray}
where \(P_k\) is the power-spectrum coefficient at the \(k\)th position in Fourier space. Note, that the mock signal \(s_i^*\) is supposed to be a real quantity, and therefore hermiticity has to be imposed in Fourier space before performing the inverse Fourier transform \citep[for details see][]{2005astro.ph..6540M}. 

Next we have to generate the additive noise contribution.
In order to draw a noise term with the correct Poissonian statistics, we first draw a random number \(N^{*}_i\) from the Poissonian distribution:
\begin{equation}
\label{eq:Poissonian}
N^{*}_i\curvearrowleft {\mathcal P}(N^{*}_i|\lambda^{*}_i) \, ,
\end{equation}
where we choose the mean observed galaxy number to be \( \lambda^{*}_i=n^{SF}_{i} \bar{\lambda}^2 \).
According to equations (\ref{eq:Observed_gal_num}) and (\ref{eq:data_model_a}) the mock noise term \(\epsilon_i^*\) can be calculated as:
\begin{equation}
\label{eq:Poissonian}
\epsilon^{*}_i= \frac{N^{*}_i-n^{SF}_{i} \bar{\lambda}^2}{\bar{\lambda}} \, .
\end{equation}
It is clear by construction that this mock noise term has vanishing mean and the correct noise covariance matrix. 
Then, according to equation (\ref{eq:data_model_a}) the mock observation is given as:
\begin{equation}
\label{eq:MOCK_DATA}
d_i^* = \sum_k \, K_{ik}\, s_k^* + \epsilon_i^* \, .
\end{equation}
The proof, that the fluctuation term \(y_i\) as generated by equation (\ref{eq:FLUCTUATION_TERM}) truly generates the correct variance is given in Appendix \ref{WIENER_VARIANCE}.

Note, that the application of the Wiener operator is a linear operation, and we can therefore rewrite equation (\ref{eq:AUGMENT_EQUATION}) as:
\begin{eqnarray}
\label{eq:UNBIASED_SIGNAL}
s_i = s^*_i + \sum_j \mat{D}^{-1}_{ij}\sum_m \sum_l \mat{K}_{mj} \mat{N}^{-1}_{ml}\left(d_l -d^*_l\right) \, ,
\end{eqnarray}
where the Wiener operator is applied to the true data \(d_i\)  and the mock observation \(d_i^*\) simultaneously. This greately reduces the CPU time required for the generation of one signal sample.

\section{Sampling the Power Spectrum}
\label{PS_SAMPLER}
As described above, the signal sampling step provides a noise-less full sky signal sample \(s_i\) consistent with the data.
The next step in the Gibbs sampling iteration scheme requires to draw power-spectrum samples from the conditional probability distribution \({\cal P}(\mat{S}|\{s_i\},\{d_i\})\).
Since in this Gibbs sampling step the perfect sky signal amplitudes \(s_i\) are known, the power-spectrum is conditionally independent of the data amplitudes \(d_i\).
Hence, in this Gibbs sampling step, we can sample the power-spectrum from the probability distribution \({\cal P}(\mat{S}|\{s_i\})\).
In the following we will show that the power-spectrum can easily be drawn from an inverse gamma distribution.

\subsection{Drawing power-spectrum samples}
\label{Power_spectrum_sampling}
According to Bayes' theorem, we can rewrite the conditional probability \({\cal P}(\mat{S}|\{s_i\})\) as:
\begin{equation}
\label{eq:COND_INDEP_PS}
{\cal P}(\mat{S}|\{s_i\}) = \frac{{\cal P}(\mat{S})}{{\cal P}(\{s_i\})} {\cal P}(\{s_i\}|\mat{S}) \, ,
\end{equation}
where \({\cal P}(\mat{S})\) is the prior for the signal covariance, \({\cal P}(\{s_i\}|\mat{S} )\) is given by equation (\ref{eq:signal_prior}) and \({\cal P}(\{s_i\})\) is a normalization constant in this Gibbs sampling step.

More specifically, we are interested in the set of matrix coefficients \(\{\mat{S}_{ij}\}\) of the covariance matrix \(\mat{S}\).
As already pointed out in section \ref{WIENER_FILTER} the signal covariance matrix of an homogeneous and isotropic universe, has an especially appealing form in Fourier space, where it takes a diagonal form.
In our application the real space covariance matrix coefficients \(\{\mat{S}_{ij}\}\) are related to their Fourier representation via the fast Fourier transform, as defined in Appendix \ref{Discrete_Fourier_transformation}. We can therefore write:
\begin{eqnarray}
\label{eq:FT_SIGNAL_COV}
\mat{S}_{ij} &=& C^2\,\sum^{N-1}_{k=0} \sum^{N-1}_{l=0}  e^{2\pi i k \frac{\sqrt{-1}}{N}}\, \hat{\hat{\mat{S}}}_{kl}\, e^{-2\pi j l \frac{\sqrt{-1}}{N}}  \nonumber \\
&=& C^2\,\sum^{N-1}_{k=0} \sum^{N-1}_{l=0}  e^{2\pi \frac{\sqrt{-1}}{N}(i\,k - j\,l )}\, \hat{\hat{\mat{S}}}_{kl} \, .
\end{eqnarray}
Then we can express the conditional distribution for the Fourier signal covariance coefficients \(\hat{\hat{\mat{S}}}_{kl}\) as:
\begin{equation}
\label{eq:COND_INDEP_PS_FT}
{\cal P}\left(\{\hat{\hat{\mat{S}}}_{kl}\}|\{s_i\}\right) = {\cal P}\left(\{\mat{S}_{ij}\}|\{s_i\}\right) \left |\frac{\partial\{\mat{S}_{ij}\}}{\partial\{\hat{\hat{\mat{S}}}_{kl}\}} \right | \, ,
\end{equation}
where 
\begin{equation}
\label{eq:JACOBIAN_DET}
\left |\frac{\partial\{\mat{S}_{ij}\}}{\partial\{\hat{\hat{\mat{S}}}_{kl}\}} \right | = \left | \rm{det}(\mat{\mathcal{J}}_{(ij)\,(kl)}) \right |\, ,
\end{equation}
is the Jacobian determinant for this coordinate transformation. As the discrete Fourier transform is proportional to a unitary matrix, this Jacobian determinant only amounts to a normalization constant, as has been demonstrated in Appendix \ref{CHANGE_TO_FFT_REPRESENTATION}.

With this definition we can rewrite the conditional probability in equation (\ref{eq:COND_INDEP_PS}), by replacing all the real space covariance matrix coefficients \(\mat{S}_{ij}\) by their Fourier representation \(\hat{\hat{\mat{S}}}_{kl}\), and normalizing it with the constant obtained from the coordinate transformation.
We can therefore write:
\begin{eqnarray}
\label{eq:COND_DENS_PS_FT}
{\cal P}(\{\hat{\hat{\mat{S}}}_{kl}\}|\{s_i\}) &=& \frac{{\cal P}(\{\hat{\hat{\mat{S}}}_{kl})\}}{{C^{N^2}\,\cal P}(\{s_i\})} {\cal P}(\{s_i\}|\{\hat{\hat{\mat{S}}}_{kl}\}) \nonumber \\
&=&\frac{{\cal P}(\{\hat{\hat{\mat{S}}}_{kl}\})}{{C^{N^2}\,\sqrt{\rm{det}\left(2\pi \hat{\hat{\mat{S}}}\right)}\,\cal P}(\{s_i\})} e^{-\frac{C^2}{2\,\hat{C}^2}\sum^{N-1}_{k=0} \sum^{N-1}_{l=0} \hat{s}^*_k \,  \hat{\hat{\mat{S}}}^{-1}_{kl} \,\hat{s}_l\,} \, , \nonumber \\
\end{eqnarray}
where we used the discrete Fourier transform definition, given in Appendix \ref{Discrete_Fourier_transformation}, to replace the real space signal amplitudes \(s_i\) by their Fourier counterparts \(\hat{s}_k\).
Introducing equation (\ref{signal_covarianceFS}) then allows us to rewrite equation (\ref{eq:COND_DENS_PS_FT}) in terms of the power-spectrum coefficients \(P_k\) as: 
\begin{equation}
\label{eq:COND_DENS_PS_COEFF_FT}
{\cal P}(\{P_k\}|\{s_i\}) =\frac{{\cal P}(\{P_k\})}{{C^{N^2}\,\cal P}(\{s_i\})} \prod_{k'=0}^{N-1} \left(2\pi\,P_{k'}\right)^{-1/2} e^{-\frac{C^2}{2\,\hat{C}^2}\sum^{N-1}_{k=0}  \frac{|\hat{s}_k|^2}{P_k} } \, , \nonumber \\
\end{equation}
where the determinant factorizes due to the diagonal form of the signal covariance matrix in Fourier space.

Note, that due to isotropy the power-spectrum is independent of direction in Fourier space, meaning the power-spectrum coefficients only depend on the modulus of the mode vector \(\vec{k}_k\):
\begin{equation}
P_k=P(\vec{k}_k)=P(|\vec{k}_k|)\, .
\end{equation}
For this reason, the angular dependence in Fourier space can be summed over.

To do so we remark that the mode vector \(\vec{k}_k\), as a geometrical object, will not change if we express it in the basis of cartesian coordinates \(\vec{k_k}=\vec{k_k}(k^1_k,k^2_k,k^3_k)\), or if we describe it in the basis of spherical coordinates \(\vec{k_k}=\vec{k_k}(|\vec{k}_k|,\varphi_k,\vartheta_k)\).
We can therefore split the multi index summation into the summation over the three spherical coordinates as:
\begin{eqnarray}
\label{eq:SUM_SPHERICAL}
\frac{C^2}{\hat{C}^2}\sum^{N-1}_{k=0}  \frac{|\hat{s}_k|^2}{P_k} &=&  \sum_{|\vec{k}_k|} \frac{1}{P(|\vec{k}_k|)}\sum_{\varphi_k} \sum_{\vartheta_k}  \frac{C^2}{\hat{C}^2} \left|\hat{s}\left(|\vec{k_k}|,\varphi_k,\vartheta_k\right)\right|^2 \nonumber \\
&=&  \sum_{|\vec{k}_k|} \frac{\sigma(|\vec{k}_k|)}{P(|\vec{k}_k|)} \nonumber \\
&=&  \sum_{m=0}^{M -1} \frac{\sigma_{m}}{P_{m}} \, ,
\end{eqnarray}
where we introduced \(\sigma(|\vec{k}_k|)=\sum_{\varphi_k} \sum_{\vartheta_k} C^2/\hat{C}^2 \left|\hat{s}\left(|\vec{k_k}|,\varphi_k,\vartheta_k\right)\right|^2\), which is the summed signal power on spherical shells around the origin in Fourier space, and the index \(m\) labels each of the \(M\) shells belonging to the different mode vector modulus \(|\vec{k}_k|\) in the Fourier box.

Several different mode vectors \(\vec{k}_k\) may have the same vector modulus \(|\vec{k}_k|\), and therefore belong to the same shell. To account for this we introduce the number \(n_m\), which counts the number of different mode vectors \(\vec{k}_k\), belonging to the  {\it \(m\)}th shell in Fourier space. This number \(n_m\), therefore counts the degrees of freedom for each of the \(M\) modes.
We can then express the product in equation (\ref{eq:COND_DENS_PS_COEFF_FT}) in terms of \(m\) as:
\begin{equation}
\label{eq:PROD_SPHERICAL}
\prod_{k=0}^{N-1} \left(2\pi\,P_k\right)^{-1/2} = \prod_{m=0}^{M-1} \left(2\pi\,P_m\right)^{-n_m/2} \, .
\end{equation}

With these definitions equation (\ref{eq:COND_DENS_PS_COEFF_FT}) turns to:
\begin{equation}
\label{eq:COND_DENS_PS_COEFF_FT_mode}
{\cal P}(\{P_k\}|\{s_i\}) =\frac{{\cal P}(\{P_k\})}{{C^{N^2}\,\cal P}(\{s_i\})} \prod_{m=0}^{M-1} \left(2\pi\,P_m\right)^{-n_m/2} e^{-\frac{1}{2}\,\frac{\sigma_{m}}{P_{m}}} \, , 
\end{equation}
When ignoring the power-spectrum prior \({\cal P}(\{P_k\})\) in the above equation (\ref{eq:COND_DENS_PS_COEFF_FT_mode}), we see that the probability distribution factorizes in the different \(P_{m}\), meaning they could be sampled independently.

If also the prior for the different \(P_{m}\) would factorize as:
\begin{equation}
\label{eq:PS_PRIOR_FACTORIZES}
{\cal P}(\{P_k\}) = \prod_{m=0}^{M-1} {\cal P}(P_m) \, ,
\end{equation}
then it is possible to sample each mode of the power-spectrum independently.

On large scales, or in the linear regime, the theory of gravitational structure formation tells us that the different Fourier modes evolve independent of each other. In these regimes the proposed power-spectrum prior would be the adequate choice.
However, we also know that nonlinear structure formation couples the different Fourier modes, the stronger the deeper we reach into the nonlinear regime. In these regimes another prior would be more adequate, but also harder to sample.

Anyhow, as already described in section \ref{LSS_SAMPLER} the entire power-spectrum sampling method requires two steps. While the different power-spectrum modes are assumed to be independent in the power-spectrum sampling step, they are not in the signal sampling step. There the different modes are coupled via the observation mask and selection function, and furthermore, the physical coupling of the different modes is represented in the data. 

Therefore, in the following we assume a power-spectrum prior, as proposed in equation (\ref{eq:PS_PRIOR_FACTORIZES}), and defer a more thorough investigation of adequate prior choices in the nonlinear regime to future work.

With this prior choice each mode can be sampled independently from the following probability density distribution:
\begin{equation}
\label{eq:COND_DENS_PS_single_mode}
{\cal P}(P_m|\{s_i\}) =\frac{{\cal P}(P_m)}{\left({C^{N^2}\,\cal P}(\{s_i\})\right)^{1/M}}  \left(2\pi\,P_m\right)^{-n_m/2} e^{-\frac{1}{2}\,\frac{\sigma_{m}}{P_{m}}} \, . 
\end{equation}
Further, we will assume a power-law behavior for the individual mode prior \({\cal P}(P_m) \propto P_m^{-\alpha}\) where \(\alpha\) is a power law index. Note, that a power-law index \(\alpha=0\) describes the flat prior, while \(\alpha=1\) amounts to the Jeffrey's prior. The Jeffrey's prior is a solution to a measure invariant scale transformation of the form \({\cal P}(P_m) dP_m = {\cal P}(\gamma\,P_m)\,\gamma dP_m\) \citep{WANDELT2004}, and therefore is a scale independent prior, as different scales have the same probability.

Inserting this power law prior in equation (\ref{eq:COND_DENS_PS_single_mode}) and imposing the correct normalization, reveals that the power-spectrum coefficients have to be sampled from an inverse gamma distribution given as:
\begin{equation}
\label{eq:INVERSE_GAMMA}
{\cal P}(P_m|\{s_i\}) =\frac{\left(\frac{\sigma_m}{2}\right)^{(\alpha-1)+n_m/2}}{\Gamma \left((\alpha-1)+\frac{n_m}{2}\right)} \frac{1}{P_m^{(\alpha+n_m/2)}} e^{- \frac{1}{2}\frac{\sigma_m}{P_m}}\, .
\end{equation}

By introducing the new variable \(x_m=\sigma_m/P_m\) and performing the change of variables we yield the \(\chi^2\)-distribution as:
\begin{equation}
\label{eq:XI_SQUARE}
{\cal P}(x_m|\{s_i\}) = \frac{x_m^{\beta_m/2-1}}{\Gamma\left(\beta_m/2\right)\,(2)^{\beta_m/2}} e^{- \frac{x_m}{2}}\, ,
\end{equation}
where \(\beta_m=2(\alpha+n_m/2-1)\).
Sampling the power-spectrum coefficients is now an easy task, as it reduces to drawing random samples from the \(\chi^2\)-distribution.
A random sample from the \(\chi^2\)-distribution for an integer \(\beta_m\) can be drawn as follows.

Let  \(z_j\) be \(\beta_m\)  independent, normally distributed random variates with zero mean and unit variance then:
\begin{equation}
\label{post:posterior_of_x_1}
x_m = \sum_{j=1}^{\beta_m} z_j^2=|\vec{z}_m|^2 \,
\end{equation}
is \(\chi^2\)-distributed, and \(\vec{z}_m\) is a \(\beta_m\) element vector, with each element being normally distributed.
The power-spectrum coefficient sample is then obtained by:
\begin{equation}
\label{eq:PS_sample}
P_m = \frac{\sigma_m}{|\vec{z}_m|^2} \, .
\end{equation}
It is easy to see that each spectrum coefficient sample is a positive quantity, this ensures that the signal covariance matrix is positive definite as it has to be by definition.

To summarize, we provide an optimal estimator for the power-spectrum coefficients, and their uncertainties.

It is also worth mentioning, that the inverse gamma distribution is a highly non-Gaussian distribution, and that for this reason, the joint estimates of signal amplitudes \(s_i\) and power-spectrum coefficients \(P_m\) are drawn from a non-Gaussian distribution.

\subsection{Blackwell-Rao estimator}
\label{Blackwell-Rao_estimator}
As described in the introduction, we seek to estimate the probability distribution \({\cal P}(\{P_m\}|\{d_i\})\), which we can now simply obtain by marginalizing over the signal samples:
\begin{eqnarray}
\label{eq:BLACKWELL_RAO}
{\cal P}(\{P_m\}|\{d_i\}) & = &\int \rm{d}\{s_i\}\,{\cal P}(\{P_m\}|\{s_i\},\{d_i\})\,{\cal P}(\{s_i\}|\{d_i\}) \nonumber \\
&= &\int \rm{d}\{s_i\}\,{\cal P}(\{P_m\}|\{s_i\})\,{\cal P}(\{s_i\}|\{d_i\}) \nonumber \\
& \approx & \frac{1}{N_{Gibbs}} \sum_{j=1}^{N_{Gibbs}} {\cal P}(\{P_m\}|\{s_i\}^j) \, , 
\end{eqnarray}
where \(\{s_i\}^j\) are the signal Gibbs samples, and \(N_{Gibbs}\) is the total number of Gibbs samples.

This result is known as the Blackwell-Rao estimator of \({\cal P}(\{P_m\}|\{d_i\})\) which is guaranteed to have a lower variance than a binned estimator \citep{WANDELT2004}.

It is worth noting, that \({\cal P}(\{P_m\}|\{s_i\})\) has a very simple analytic form, and therefore equation (\ref{eq:BLACKWELL_RAO}) provides an analytic approximation to \({\cal P}(\{P_m\}|\{d_i\})\) based on the Gibbs samples.
All the information on \({\cal P}(\{P_m\}|\{d_i\})\) is therefore contained in the \(\sigma_m\) of the individual Gibbs steps, which generate a data set of size \(\O(m_{max}\,N_{Gibbs})\), where \(m_{max}\) is the maximal number of independent modes.
In addition, to being a faithful representation of \({\cal P}(\{P_m\}|\{d_i\})\) the Blackwell-Rao estimator is also a computationally efficient representation, which allows to calculate any moment of \({\cal P}(\{P_m\}|\{d_i\})\) as:
\begin{equation}
\label{eq:EST_HIGHER_MOMENTS}
\langle P_m\,P_{m'} ... P_{m''}\rangle|_{{\cal P}(P_m|\{d_i\})} \approx \frac{1}{N_{Gibbs}} \sum_{j=1}^{N_{Gibbs}} \langle P_m\,P_{m'} ... P_{m''}\rangle|_{{\cal P}(P_m|\{s_i\}^j)} \, ,
\end{equation}
where each of the terms on the right handside can be calculated analytically.

For the inverse gamma distribution given in equation (\ref{eq:INVERSE_GAMMA}) we can then simply calculate the mean of the probability distribution \({\cal P}(P_m|\{d_i\})\) as:
\begin{equation}
\label{eq:BR_MEAN}
\langle P_m \rangle|_{{\cal P}(P_m|\{d_i\})} \approx  \frac{ \frac{1}{N_{Gibbs}} \sum_{j=1}^{N_{Gibbs}} \sigma_m^j}{2(\alpha-2)+n_m}   \, ,
\end{equation}
and in analogy the variance as:
\begin{equation}
\label{eq:BR_VARIANCE}
\left \langle \left [ P_m -\langle P_m \rangle\right]^2 \right \rangle|_{{\cal P}(P_m|\{d_i\})} \approx  \frac{\frac{1}{N_{Gibbs}}\,\sum_{j=1}^{N_{Gibbs}}\left(\sigma_m^j\right)^2}{4\,\left((\alpha-2)+n_m/2\right)^2\left((\alpha-3)+n_m/2\right)} \, .
\end{equation}
The Blackwell-Rao estimator also allows us to demonstrate another remarkable property of the Gibbs sampling approach. Although a specific power-spectrum prior has to be employed during the Gibbs analysis of the data, a post processing analysis of the power-spectrum can be performed with any desired power-spectrum prior.
Lets assume one prefers to perform a post processing analysis with a power-spectrum prior \({\cal P'}(\{P_m\})\), rather than with the prior \({\cal P}(\{P_m\})\), which was employed during Gibbs sampling. We therefore want to estimate the power-spectrum from the following posterior:
\begin{eqnarray}
\label{eq:BLACKWELL_RAO}
{\cal P'}(\{P_m\}|\{d_i\}) & = & {\cal P'}(\{P_m\}) \frac{{\cal P}(\{d_i\}|\{P_m\})}{{\cal P}(\{d_i\})}\nonumber \\ 
& = & \frac{{\cal P'}(\{P_m\})}{{\cal P}(\{P_m\})}\, {\cal P}(\{P_m\}|\{d_i\})\nonumber \\ 
&= & \frac{{\cal P'}(\{P_m\})}{{\cal P}(\{P_m\})} \int \rm{d}\{s_i\}\,{\cal P}(\{P_m\}|\{s_i\})\,{\cal P}(\{s_i\}|\{d_i\}) \nonumber \\
&= & \int \rm{d}\{s_i\}\,{\cal P'}(\{P_m\})\,\frac{{\cal P}(\{s_i\}|\{P_m\})}{{\cal P}(\{s_i\})}\,{\cal P}(\{s_i\}|\{d_i\}) \nonumber \\
& \approx & \frac{1}{N_{Gibbs}} \sum_{j=1}^{N_{Gibbs}} {\cal P'}(\{P_m\})\,\frac{{\cal P}(\{s_i\}^j|\{P_m\})}{{\cal P}(\{s_i\}^j)} \, , 
\end{eqnarray}
where we simply made use of the Bayes theorem. Since \({\cal P}(\{s_i\}|\{P_m\})\) is a simple Gaussian distribution, and therefore given analytically, the posterior \({\cal P'}(\{P_m\}|\{d_i\})\) can be calculated with any desired power-spectrum prior in a post-processing step.

\section{The prior and the cosmic variance}
\label{Prior_and_Variance}
The Gibbs sampling procedure consists of the two basic steps, of first sampling perfect noise-less full sky signal samples \(s_i\) and then sampling the power-spectrum coefficients \(P_m\) given \(s_i\).

Therefore, the probability distribution \({\cal P}(P_m|\{s_i\})\), given in equation (\ref{eq:INVERSE_GAMMA}), encodes our knowledge on the power-spectrum coefficients \(P_m\), if we had perfect knowledge of the true signal amplitudes \(s_i\).

It is clear, that in the case of perfect observations the full posterior distribution for the power-spectrum coefficients \({\cal P}(P_m|\{d_i\})\) would reduce to that one given in equation (\ref{eq:INVERSE_GAMMA}).

This is, because in the case of perfect full-sky and noise-less observations, the signal posterior would collapse to a Dirac delta distribution, due to the vanishing noise covariance matrix.
This means, that in this case the signal amplitudes \(s_i\) can be estimated with zero variance.

However, measuring the \(P_m\) to arbitrary precision will never be possible. The power-spectrum coefficients depend on the data through the \(\sigma_m\), which measure the actual fluctuation power in the observed Universe. It is clear that the probability distribution function (\ref{eq:INVERSE_GAMMA}) for the \(P_m\) will not reduce to a Dirac delta distribution, even though the \(\sigma_m\) have been measured perfectly.

Owing to this fact, there will always remain some uncertainty in the power-spectrum estimation, even in the case of perfect measurements.
This residual uncertainty is well known as cosmic variance, which is the direct consequence of only observing just one specific matter field realization. 

The Gibbs sampling approach, as proposed here, takes this cosmic variance into account, by drawing samples from the probability distribution \({\cal P}(P_m|\{s_i\})\), which obey the correct statistical properties.

%\begin{figure*}
%	\centering
%	{
%	\begin{picture}(100,180)

%	\put(-210,100){\rotatebox{90}{\(F_i\)}}
%	\put(-135,15){\(r_i\) [Mpc/h]}

%	\put(-30,100){\rotatebox{90}{\(ra \left[^{\circ}\right]\)}}
%	\put(135,5){\(dec \left[^{\circ}\right]\)}

 % 	\put(-200,180){
%	\rotatebox{270}
%	{
%		\includegraphics[bb = 66 169 546 657,width=0.305\textwidth,clip=true]{figs/selfunc.eps}
%	}
%	}

%	\put(-30,14){
%	{
%		\includegraphics[bb =44 -6 381 165 ,width=0.647\textwidth,clip=true]{figs/SKYMASK.eps}
%	}
%	}
%\end{picture}
%}
%\caption{Selection function and two dimensional sky mask.}
%	\label{fig:TEST_SEL_WIN}
%\end{figure*}

\begin{figure*}
\centering{\resizebox{1.\hsize}{!}{\includegraphics{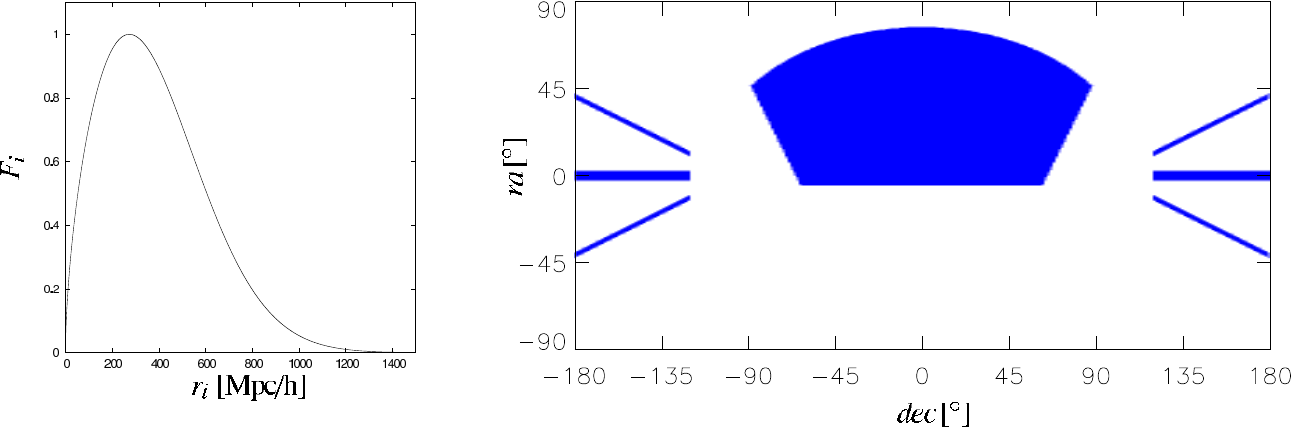}}}
\caption{Selection function and two dimensional sky mask.}
\label{fig:TEST_SEL_WIN}
\end{figure*}

\subsection{Flat versus Jeffrey's prior}
The main characteristics of \({\cal P}(P_m|\{s_i\})\) can be summarized in terms of the mean, mode and variance, which allows us to discuss the influence of the actual power law prior choice. 
The mode of the inverse gamma distribution (\ref{eq:INVERSE_GAMMA}) is the most frequently used estimator for the power-spectrum coefficients when using fast Fourier transform techniques. The mode \(P_m^*\) is defined as \(P_m^* \in \{P_m | \forall P_l : \, \, {\cal P}(P_l|\{s_i\}) \le {\cal P}(P_m|\{s_i\}) \}  \), which is simply the value of \(P_m\) which maximizes the distribution. For the inverse gamma distribution it is given as:
\begin{equation}
\label{eq:mean_inverse_Gamma}
P_m^* = \frac{\sigma_m}{(2\alpha+n_m)} \, .
\end{equation}
Assuming  a flat prior \(\alpha=0\), this immediately returns the frequently used and simple power-spectrum estimator:
\begin{equation}
\label{eq:simple_power_spec_est}
P_m^* = \frac{\sigma_m}{n_m} \, ,
\end{equation}
\citep[see e.g.][]{Volker2008}.

However, note, that a flat prior is a questionable choice, when measuring a variance, which is a scale parameter, as it does not correspond to maximal ignorance but biases towards large excursions from zero \citep{WANDELT2004}.
Additionally the flat prior does not permit to sample every mode in the three dimensional Fourier box with finite variance, which can be easily seen by looking at the variance of the inverse gamma distribution given as:
\begin{eqnarray}
\label{eq:variance_inverse_Gamma}
\left \langle \left [ P_m -\langle P_m \rangle\right]^2 \right \rangle  = \frac{\sigma_m^2}{4\,( (\alpha-2)+n_m/2)^2((\alpha-3)+n_m/2)}
\end{eqnarray}
which is only finite for \(2\alpha+n_m>6\).

In a three dimensional cubic Fourier box the minimal number \(n_m\) for a mode is \(min(n_m)=6\) (except for the zero mode). This corresponds to the six mode vectors \(\vec{k_k}\) with same vector modulus \(|\vec{k_k}|\) along the three axes in Fourier space. Nevertheless, a flat prior (\(\alpha=0\)) requires \(n_m>6\) in order to sample the modes with finite variance, which cannot be fulfilled for these modes.

Therefore, we favor the Jeffrey's prior with \(\alpha=1\), which requires only \(n_m>4\) to sample each mode, except for the zero mode, with finite variance. Jeffrey's prior is also scale invariant, and therefore does not introduce any bias on a log-scale.

\subsection{Informative prior}
\label{Informative_Prior}
The prior discussed in the previous section is a maximal ignorance prior in the sense, that every scale has the same probability. This prior therefore allows for large excursions around the true value of the power-spectrum. This is especially important when sampling the largest scales in a galaxy survey, which are poorly constrained by measurements. A maximum ignorance prior will therefore require to sample a huge space of possible power-spectrum configurations.

However, one can argue, that knowledge about the largest scales exists, either through theory, or CMB measurements, which provide detailed information on the largest scales.

For this reason, it might be interesting to incorporate this a priori knowledge on the power-spectrum into our sampling scheme, and therefore allowing for a more efficient strategy to sample the space of possible power-spectrum configurations.

The most simple informative prior can be obtained by limiting the range of the Jeffrey's prior, by setting the Jeffrey's prior equal to zero for power-spectrum excursion of more than a certain factor:
\begin{eqnarray}
\label{eq:INF_JEFFREYS_PRIOR}
{\cal P}(P_m) &\propto& \left \{ \begin{array}{ll}
  P_m^{-\alpha} & \quad \mbox{for $ P^{Prior}_m / \tau \le P_m \le P^{Prior}_m\, \tau $}\\
  0 & \quad \mbox{otherwise}\\ \end{array} \right. 
\end{eqnarray}
where \(P^{Prior}_m\) is our best guess power-spectrum prior, and \(\tau\) is a factor, which permits a certain range around the prior power-spectrum.
The sampling scheme, which arises by implementing this prior, is basically the same as described in section \ref{Power_spectrum_sampling}, with the only exception that power specrum coefficients \(P_m\), which do not fulfill the requirement described in equation (\ref{eq:INF_JEFFREYS_PRIOR}), are rejected and have to be resampled.
This prior would still be a maximum ignorance prior over the allowed range.

Another possible informative prior, which allows to sample the entire range, but favoring the region, in which we expect the true power-spectrum to exist, can be very easily found by assuming some a priori knowledge on the \(\sigma_m\). For example, this can be achieved by incorporating an independent observation to the sampling scheme.
In this case we can again assume an inverse gamma distribution for the power-spectrum prior:
\begin{equation}
\label{eq:INVERSE_GAMMA_PRIOR}
{\cal P}(P_m) =\frac{\left(\frac{\sigma^{Prior}_m}{2}\right)^{(\alpha^{Prior}-1)+n_m^{Prior}/2}}{\Gamma \left((\alpha^{Prior}-1)+\frac{n_m^{Prior}}{2}\right)} \frac{1}{P_m^{(\alpha^{Prior}+n_m^{Prior}/2)}} e^{- \frac{1}{2}\frac{\sigma_m^{Prior}}{P_m}}\, ,
\end{equation}
where \(\sigma^{Prior}_m\) describes our a priori knowledge on the \(\sigma_m\), \(\alpha^{Prior}\) is the spectral index of of our power law prior, which we choose \(\alpha^{Prior}=1\) to be the Jeffrey's prior, and \(n_m^{Prior}\) is the number of mode counts for our prior guess.
Note, that the combination of \(\alpha^{Prior}\) and \(n_m^{Prior}\) defines how sharp this prior would be.
As we want our prior to contain as little information as possible, we choose \(n_m^{Prior}=5\) as this is the minimal number of modes, which lead to a finite variance with the Jeffrey's prior.

Introducing an inverse gamma prior will then yield again a inverse gamma distribution for the power-spectrum sampling procedure: 
\begin{equation}
\label{eq:INVERSE_GAMMA_WITH_PRIOR}
{\cal P}(P_m|\{s_i\}) =\frac{\left(\frac{\sigma_m^{Prior}+\sigma_m}{2}\right)^{(\alpha^{Prior}-1)+(n_m^{Prior}+n_m)/2}}{\Gamma \left((\alpha^{Prior}-1)+\frac{(n_m^{Prior}+n_m)}{2}\right)} \frac{e^{- \frac{1}{2}\frac{(\sigma_m^{Prior}+\sigma_m)}{P_m}}}{P_m^{(\alpha^{Prior}+(n_m^{Prior}+n_m)/2)}}  \, .
\end{equation}
By introducing \(x_m=(\sigma_m^{Prior}+\sigma_m)/P_m\), this can again be rewritten as a \(\chi^2\)-distribution:
\begin{equation}
\label{eq:XI_SQUARE_with_prior}
{\cal P}(x_m|\{s_i\}) = \frac{x_m^{\beta_m/2-1}}{\Gamma\left(\beta_m/2\right)\,(2)^{\beta_m/2}} e^{- \frac{x_m}{2}}\, ,
\end{equation}
where \(\beta_m=2(\alpha^{Prior}+(n_m^{Prior}+n_m)/2-1)\).

A power-spectrum sample is then obtained in the same fashion as described in section \ref{Power_spectrum_sampling}, by
\begin{equation}
\label{eq:PS_sample}
P_m = \frac{\sigma_m^{Prior}+\sigma_m}{|\vec{z}_m|^2} \, ,
\end{equation}
with \(\vec{z}_m\) being \(\beta_m\) element vector, with each element being normally distributed with zero mean and unit variance.

As an example of incorporating theoretical information one can generate the \(\sigma_m^{Prior}\) by:
\begin{equation}
\label{eq:PS_Prior}
\sigma_m^{Prior}  = n_m^{Prior}\, P_m \, ,
\end{equation}
which will yield on average the prior power-spectrum.
In this fashion the Gibbs sampler will sample around our prior guess for the power-spectrum.

Note, that this method provides a possible interface for joint power-spectrum estimation from a joint CMB  and large scale structure analysis, where the \(\sigma_m^{Prior}=\sigma_m^{CMB}\) will be obtained from a CMB analysis step.

\subsection{Hidden prior}
\label{hidden_prior}
In the previous section, we described how to sample each mode \(|\vec{k}_k|\) of the Fourier box individually, which yields extremely fine frequency resolution in the power-spectrum estimate. As in practice such a high frequency resolution is not required, or desired, one allows each shell \(m'\) to have a finite thickness \(\Delta k_m'\), rather than treating it as infinitely thin.

As the newly designated shells \(m'\) have a finite thickness, different infinitely thin shells \(m\) now belong to the same shell \(m'\).

With the shells having a finite thickness, different close by Fourier modes \(|\vec{k}_k|\) fall into the same bin, and therefore an assumption about the functional shape \(f_m(|\vec{k_k}|)\), which yields the correct weighting, over this shell \(m'\) around the central mode \(|\vec{k}|_{m'}\) must be assumed:
\begin{equation}
\label{eq:PS_sample_piece_function}
P_m = A_{m'} f^{m'}_m \mbox{     for \( \vec{k_k} \in \left[|\vec{k}|_m-\Delta k_m/2,|\vec{k}|_m\Delta k_m/2\right]\) } \, ,
\end{equation}
where \(A_{m'}\) is the constant amplitude for the {\it\(m'\)th} shell. Usually, the shape of the power-spectrum \(f^{m'}_m\) is assumed to be constant over the shell width  \(\Delta k_m'\), which amounts to "binning" the power-spectrum. However, in general \(f^{m'}_m\) could assume any desired functional shape.

With this assumption, the estimate of the power-spectrum is further constrained by the assumed functional shape \(f^{m'}_m\), and we seek to estimate the set of power-spectrum coefficients \(\{P_m\}\) from a probability distribution given as:
\begin{eqnarray}
\label{1st:cont_spec_posterior_binning}
{\cal P}(\{P_m\}|\{s_i\},\{f^{m'}_m\}) &=&\frac{{\cal P}(\{f^{m'}_m\}|\{P_m\})}{{\cal P}(\{f^{m'}_m\}|\{s_i\})}  \, \frac{{\cal P}(\{P_m\}) {\cal P}(\{s_i\}|\{P_m\},\{f^{m'}_m\})}{{\cal P}(\{s_i\})}  \nonumber \\
&=&\frac{{\cal P}(\{f^{m'}_m\}|\{P_m\})}{{\cal P}(\{f^{m'}_m\}|\{s_i\})}  \, \frac{{\cal P}(\{P_m\}) {\cal P}(\{s_i\}|\{P_m\})}{{\cal P}(\{s_i\})}  \, , \nonumber \\
\end{eqnarray}
where we assumed conditional independence \({\cal P}(\{s_i\}|\{P_m\},\{f^{m'}_m\})={\cal P}(\{s_i\}|\{P_m\})\) of the functional shape, once all power-spectrum coefficients are given.
%\begin{figure*}
%	\centering
%	{
%	\begin{picture}(100,240)

%	\put(-205,115){\rotatebox{90}{\(y\) [Mpc/h]}}
%	\put(-100,10){\(x\) [Mpc/h]}
	
%	\put(60,115){\rotatebox{90}{\(z\) [Mpc/h]}}
%	\put(165,10){\(x\) [Mpc/h]}

 % 	\put(-200,15){
%	\rotatebox{0}
%	{
%		\includegraphics[bb =0 0 720 720,width=0.45\textwidth,clip=true]{figs/DATEN1.eps}
%	}
%	}

%	\put(65,15){
%	{
%		\includegraphics[bb =0 0 720 720,width=0.45\textwidth,clip=true]{figs/DATEN2.eps}
%	}
%	}
%\end{picture}
%}
%\caption{Volume rendering of observed galaxy densities in two different projections.}
%	\label{fig:volume_rendering_observation}
%\end{figure*}

\begin{figure*}
\centering{\resizebox{1.\hsize}{!}{\includegraphics{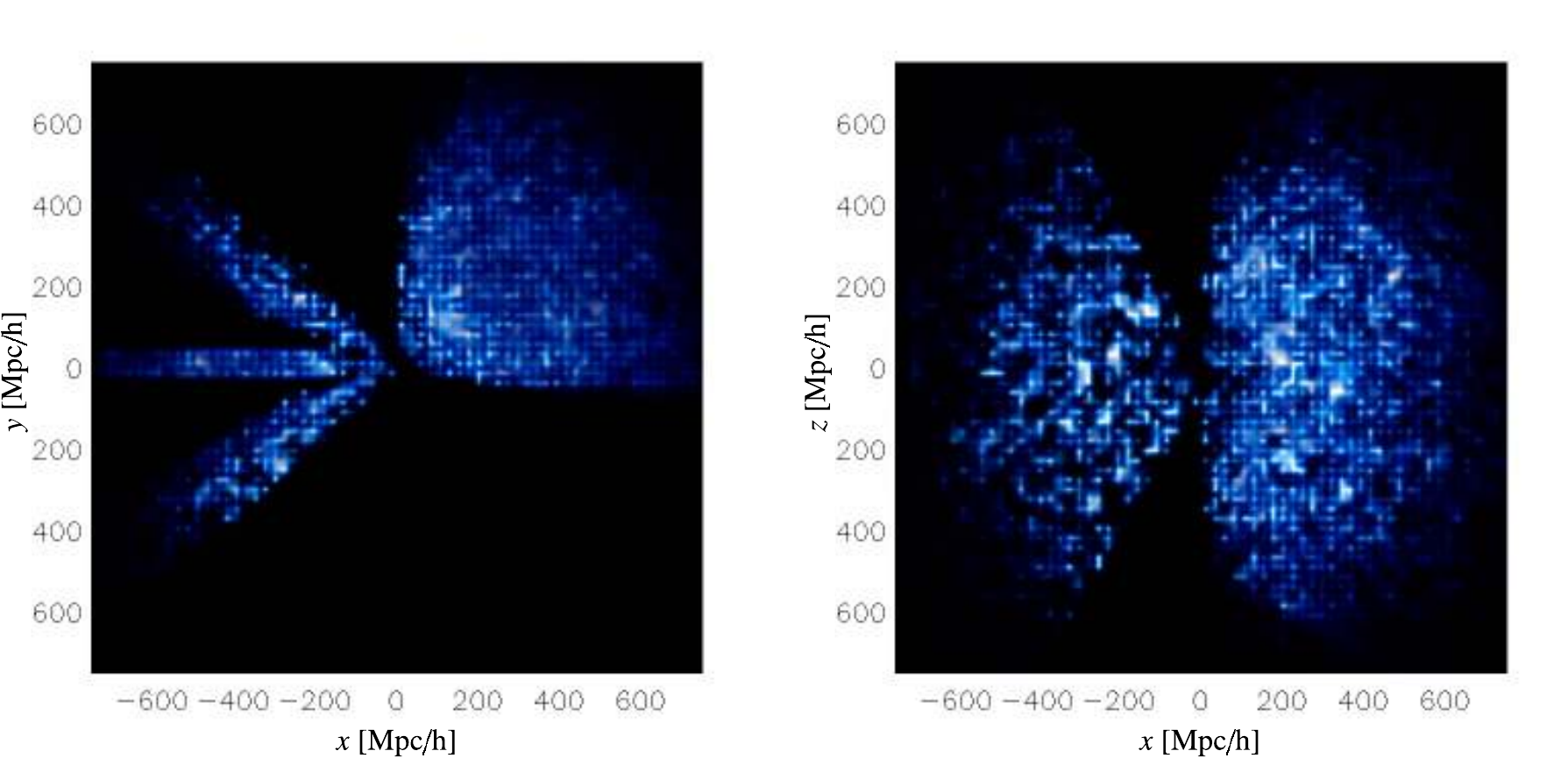}}}
\caption{Volume rendering of observed galaxy densities in two different projections.}
\label{fig:volume_rendering_observation}
\end{figure*}

The usual implicit assumption when introducing this kind of power-spectrum binning is:
\begin{equation}
\label{eq:Pimplicit assumption}
{\cal P}(\{f^{m'}_m\}|\{P_m\})¸\propto \prod_m \delta^D(A_m'\,f^{m'}_m-P_m)\, ,
\end{equation}
meaning, we claim to know the exact functional shape of the power-spectrum within the shells \(m'\).

This reduces the amount of free parameters to be sampled, from the set of \(N_m\) power-spectrum coefficients \(P_m\) to the set of \(N_{m'}\) amplitudes \(A_{m'}\). 

If, for instance, we knew the exact shape of the entire power-spectrum, sampling the power-spectrum could be reduced to the task of just sampling the overall amplitude. Such an approach is persued, when trying to sample the cosmological parameters.

However, with the above definition, sampling the power-spectrum reduces to sample the amplitudes \(A_{m'}\), and we can write:
\begin{eqnarray}
\label{1st:cont_spec_posterior_binning}
{\cal P}(\{A_{m'}\}|\{s_i\},\{f^{m'}_m\}) &=& {\cal P}(\{P_m\}|\{s_i\},\{f^{m'}_m\}) \left| \frac{\partial \{P_m\} }{\partial \{A_{m'}\}} \right|  \nonumber \\
&=& \prod_m \frac{\left(\frac{\sigma_m}{2\,f^{m'}_m}\right)^{(\alpha-1)+n_m/2}}{\Gamma \left((\alpha-1)+\frac{n_m}{2}\right)} \frac{ e^{- \frac{1}{2}\frac{\sigma_m}{A_{m'}\,f^{m'}_m}}}{(A_{m'})^{(\alpha+n_m/2)}}\nonumber \\
&=& \prod_{m'}\prod_{m\,\in \, m'} \frac{\left(\frac{\sigma_m}{2\,f^{m'}_m}\right)^{(\alpha-1)+n_m/2}}{\Gamma \left((\alpha-1)+\frac{n_m}{2}\right)} \frac{ e^{- \frac{1}{2}\frac{\sigma_m}{A_{m'}\,f^{m'}_m}}}{(A_{m'})^{(\alpha+n_m/2)}}\nonumber \\
&=& \prod_{m'}\frac{\left(\frac{\sigma_{m'}}{2}\right)^{(\alpha-1)+n_{m'}/2}}{\Gamma \left((\alpha-1)+\frac{n_{m'}}{2}\right)} \frac{ e^{- \frac{1}{2\,A_{m'}} \sum_{m\, \in \, m'}\frac{\sigma_m}{\,f^{m'}_m}}}{(A_{m'})^{(\alpha+n_{m'}/2)}}\nonumber \\
&=& \prod_{m'}\frac{\left(\frac{\sigma_{m'}}{2}\right)^{(\alpha-1)+n_{m'}/2}}{\Gamma \left((\alpha-1)+\frac{n_{m'}}{2}\right)} \frac{ e^{- \frac{1}{2} \frac{\sigma_{m'}}{A_{m'}}}} {(A_{m'})^{(\alpha+n_{m'}/2)}}\, , \nonumber \\
\end{eqnarray}
where \(\sigma_{m'}=\sum_{m\, \in \, m'}\sigma_m/f^{m'}_m\) and \(n_{m'}= \sum_{m\, \in \, m'} n_{m}\).
As this probability distribution factorizes in the amplitudes \(A_{m'}\), each of these amplitudes can be independently sampled from the inverse gamma distribution, with the method described in section \ref{Power_spectrum_sampling}.

A power-spectrum coefficient \(P_m\) is therefore obtained as:
\begin{equation}
\label{1st:binned_PM_estimate}
P_m = \frac{\sigma_{m'}\,f^{m'}_m}{|\vec{z}_{m'}|^2} \, ,
\end{equation}
where \(\vec{z}_{m'}\) is a \(n_{m'}\) component vector, with the elements being Gaussian distributed with zero mean and unit variance.

Note, since \(n_{m'}\ge n_{m}\) the variance for the power-spectrum coefficients \(P_m\) is reduced. This is the result of reducing the amount of free parameters, by introducing "binning", as now each amplitude estimate for the \(A_{m'}\) is based on more supporting points than the individual \(P_m\).

Due to the finite shell width \(\Delta k_{m'}\) neighboring modes are coupled. This fact could be exploited to circumvent the problem of missing mode coupling in the nonlinear regime. For example, if the different shells would be logarithmically spaced, it is possible to sample the largest scales independently, while towards the non-linear regime, the modes get more and more coupled. From a physical point of view, the logarithmic spacing would therefore be best suited for this problem. 
On the other hand, introducing "binning" to the power-spectrum sampling procedure, makes the method insensitive to fluctuations an scales smaller than the shell width \(\Delta k_{m'}\), and a \(\Delta k_{m'}\) should therefore be chosen carefully in order not to miss features we intend to recover.

It is also important to note, that the variance in the power-spectrum sampling step defines the stepsize for the random walk, to sample the joint probability distribution of signal and power-spectrum. If the "binning" is unreasonable large, and therefore variance is dramatically reduced, it takes much longer to explore the joint space of signal amplitudes \(s_i\) and power-spectrum coefficients \(P_m\). 
For this reason, we prefer to sample with rather high spectral resolution for the power-spectrum.

%\begin{figure*}
%	\centering
%	{
%	\begin{picture}(100,240)

%	\put(-205,130){\rotatebox{90}{\(\xi_l^i\)}}
%	\put(-90,0){\(k\) [h/Mpc]}
	
%	\put(60,130){\rotatebox{90}{\(C_l(n)\)}}
%	\put(190,0){\(n\)}

 % 	\put(-200,15){
%	\rotatebox{0}
%	{
%		\includegraphics[bb =104 78 433 395,width=0.45\textwidth,clip=true]{figs/BIT_MI_PRIOR.eps}
%	}
%	}

%	\put(65,13.5){
%			\rotatebox{0}
%	{
%		\includegraphics[bb =90 76 396 363,width=0.465\textwidth,clip=true]{figs/MI_PRIOR/CORR_COEFF_MI_PRIOR.eps}
%	}

%	}
%\end{picture}
%}
%\caption{Test results from low resolution test run. The left panel shows the results of the Burn-in test for eight sequential \(\xi_l^i\) as a function of Fourier modes \(k\). The right panel displays results for the correlation coefficients \(C_l(n)\) of the correlation length test for all Fourier modes.}
%	\label{fig:BIT_CORRCOEFF}
%\end{figure*}
\begin{figure*}
\centering{\resizebox{1.\hsize}{!}{\includegraphics{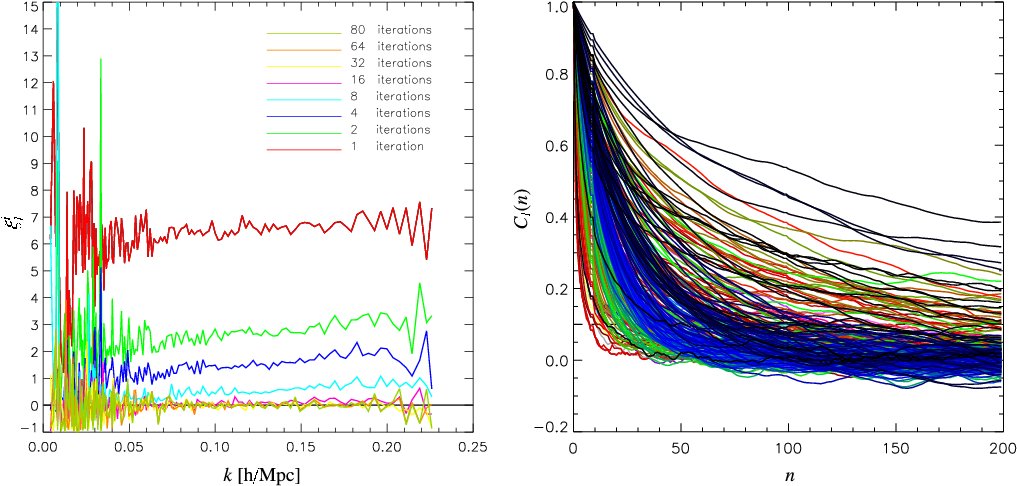}}}
\caption{Test results from low resolution test run. The left panel shows the results of the Burn-in test for eight sequential \(\xi_l^i\) as a function of Fourier modes \(k\). The right panel displays results for the correlation coefficients \(C_l(n)\) of the correlation length test for all Fourier modes.}
\label{fig:BIT_CORRCOEFF}
\end{figure*}

\section{Numerical Implementation}
\label{NUMERICAL_IMPLEMENTATION}
Our numerical implementation of the large scale structure Gibbs sampler is called \textsc{ARES} (Algorithm for REconstruction and Sampling).
It can be separated into the described two Gibbs sampling steps of estimating a signal sample, which involves solving large systems of equations, and sampling the power-spectrum, by drawing random samples from the inverse gamma distribution.

\subsection{Inversion of matrices}
The signal sampling step is by far the most numerically demanding step as it requires fast and efficient inversions of large matrices. \textsc{ARES} utilizes the fast operator based conjugate gradients inversion technique as presented in the \textsc{ARGO} code \citep[][]{Kitaura}, which has recently been applied to Sloan Digital Sky Data, to obtain matter field reconstructions \citep[][]{KITAURA2009}.\nextext{ Operator based inversion techniques have been previously developed for CMB data analysis \citep[][]{WANDELT2004,2004ApJS..155..227E,JEWELL2004}.} 
 
Rather than  requiring to store the matrices under consideration explicitely in computer memory, which would be numerically prohibitive, this approach only requires to know how these matrices would act on a vector, and therefore reduces the required amount of computer memory to numerically feasible amounts.
Further, it is possible to reduce the scaling of the most expensive matrix operation to that one of a fast Fourier transform. \textsc{ARES} uses the FFTW3 library, which therefore reduces the scaling of the most expensive operation to \(\mathcal{O}(N\,log(N))\) \citep{FFTW05}. 

The FFTW3 library also incorporates the feature of parallel Fourier transforms, which allows for straight forward parallelization of our code.

%\begin{figure*}
%	\centering
%	{
%	\begin{picture}(160,160)
%	\put(0,70){\rotatebox{90}{walltime [s]}}
%	\put(70,0){Gibbs iteration}
%	\put(160,0){\(\)}
%	\put(5,10){
%%	{
%		\includegraphics[bb = 101 78 437 395,width=0.3\textwidth,clip=true]{figs/walltime.eps}
%	}
%	}
%	\end{picture}
%	}
%	\caption{Wall clock times for a single map making step over 10000 Gibbs iterations. The red line denotes the average wall clock time for one matrix inversion step.}
%	\label{fig:WALL_TIMES}
%\end{figure*}

\begin{figure*}
\centering{\resizebox{0.4\hsize}{!}{\includegraphics{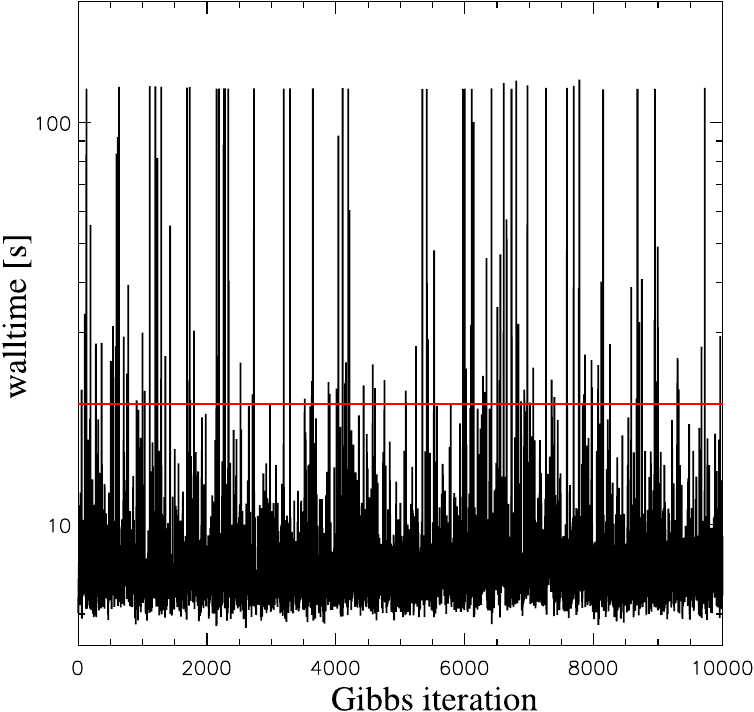}}}
\caption{Wall clock times for a single map making step over 10000 Gibbs iterations. The red line denotes the average wall clock time for one matrix inversion step.}
\label{fig:WALL_TIMES}
\end{figure*}

\subsection{Random number generation}
Our random number generation relies on a pseudo random number generator as provided by the GNU scientific library (gsl) \citep{GSL}.
In particular, we use the Mersenne Twister MT19937, with 32-bit word length, as provided by the gsl\_rng\_mt19937 routine.

The Mersenne Twister algorithm was designed for Monte Carlo simulations, where primarily good quality numbers and speed are decisive \citep{MERSENNE_TWISTER}.
It has been proven to have a period of \(2^{19937}-1\), where for our application in practice we usually require only about \(2^{35}\) unique random numbers. Also note, that the very high order of dimensional equidistribution guarantees negligible serial correlation in the output sequence.
The Mersenne Twister algorithm passed several tests for statistical randomness, including the Diehard tests.

\subsection{Parallelization}
\label{Parallelization}
Parallelization of the code is a crucial issue, since CPU time is the main limiting factor of our method.
Even though the conjugate gradient method allows for very efficient matrix inversions, for a complete data analysis one has to perform many of these matrix inversions, and therefore the total prefactor of the algorithm increases.

In figure \ref{fig:WALL_TIMES} we show a typical progression for the wall clock times of the map making algorithm during the Gibbs sampling chain for a low resolution simulation with \(N_{VOX}=64^3\) and a setup as described in section \ref{Gaussian_Test_Cases}.

As can be seen, the wall clock times may vary from Gibbs iteration to Gibbs iteration.

This variation is due to complexity of the actual problem in the Gibbs iteration, which may require more conjugate gradients steps to reach the desired numerical accuracy.

The average creation time for one Gibbs sample in this low resolution simulation is about \(20\) seconds. Where this test was run on a single Desktop CPU. Doubling the resolution will require roughly eight times longer to create a sample. Hence, a \(N_{VOX}=128^3\) simulation will require roughly \(5\) minutes and a \(N_{VOX}=512^3\) about \(8\) hours to create a single Gibbs sample.

This immediately clarifies the need to parallelize the code, as usually tenth of thousands of Gibbs samples are required.

There are in principle two different approaches to parallelize the code.

The most demanding step in the Gibbs sampling chain is the map making process. One could therefore parallelize the map making algorithm, which in principle requires to parallelize the fast Fourier transform. The FFTW3 library provides parallelized fast Fourier transform procedures, and implementation of those is straight forward \citep{FFTW05}.
However, optimal speed up cannot be achieved.
The other approach relies on the fact that our method is a Monte Carlo process, and each CPU can therefore calculate its own Markov chain.
In this fashion we gain optimal speed up and the possibility to initialize each chain with different initial guesses. 

The major difference between these two parallelization approaches is, that with the first method one tries to calculate a rather long sampling chain, while the latter one produces many shorter chains.

As we will see in the next section, successive Gibbs samples are highly correlated in the low signal to noise regime and producing a larger number of independent samples requires longer chains.
This problem, however, can be partially overcome by initializing each Markov chain with an independent power-spectrum guess.

%\begin{figure*}
%	\centering
%	{
%	\begin{picture}(100,240)

%	\put(-205,86){\rotatebox{90}{Probability density}}
%	\put(-100,0){\(P(k)\)}
	
%	\put(60,106){\rotatebox{90}{\(k\) [h/Mpc]}}
%	\put(316,116){\rotatebox{90}{\(\mat{C}_{l\,l'}\)}}
%	\put(175,0){\(k\) [h/Mpc]}

 % 	\put(-200,15){
%		\rotatebox{0}
%	{
%		\includegraphics[bb =83 74 398 384,width=0.4\textwidth,clip=true]{figs/INF_PRIOR/modestat_INF_PRIO.eps}
%	}
%	}

%	\put(65,10){
%			\rotatebox{0}
%	{
%			\includegraphics[bb =42 9 424 329,width=0.483\textwidth,clip=true]{figs/MI_PRIOR/CORRMAT_MIPRIOR.eps}
%	}
%	}
%\end{picture}
%}
%\caption{The left panel shows the mode statistics for some selected Fourier modes. The green vertical line denotes the true underlying power-spectrum. The red line denotes the averaged mean obtained from the Gibbs chain, and the black vertical line denotes the power-spectrum of the specific matter field realization. The right panel shows the power-spectrum covariance matrix estimated from the 40,000 Gibbs samples.}
%	\label{fig:MODE_STAT_CORR_MAT_MI_PRIOR}
%\end{figure*}

\begin{figure*}
\centering{\resizebox{1.\hsize}{!}{\includegraphics{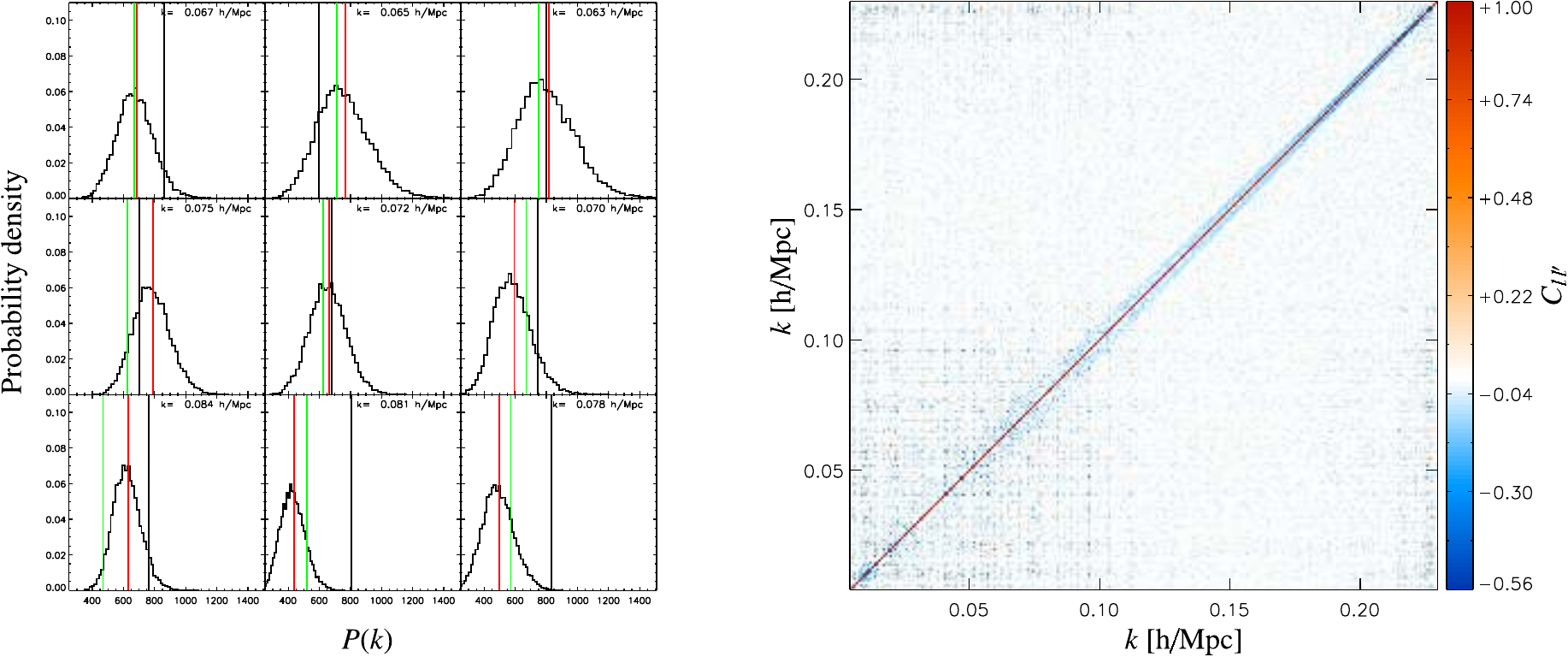}}}
\caption{The left panel shows the mode statistics for some selected Fourier modes. The green vertical line denotes the true underlying power-spectrum. The red line denotes the averaged mean obtained from the Gibbs chain, and the black vertical line denotes the power-spectrum of the specific matter field realization. The right panel shows the power-spectrum covariance matrix estimated from the 40,000 Gibbs samples.}
\label{fig:MODE_STAT_CORR_MAT_MI_PRIOR}
\end{figure*}

\section{Testing \textsc{ARES}}
\label{Gaussian_Test_Cases}
In this section, we apply \textsc{ARES} to simulated mock observations, where the underlying dark matter signal is perfectly known. With these mock observations we will be able to demonstrate that the code produces results consistent with the theoretical expectation. In addition, we will gain insight in how the code may perform in real-world applications, when CPU time is limited.
Therefore, we will set up Gaussian mock cases, designed to highlight some specific feature of the code.

\subsection{Setting up a Gaussian Mock observation}
\label{SET_UP_MOCK_OBS}
For the cases studied in this section we set up a set of low resolution mock observations based on Gaussian random fields, which will allow us to calculate a large number of samples in reasonable computational times.

These mock observations are generated on a three dimensional cartesian box with \(N_{side}=64\), corresponding to \(N_{vox}=262144\) volume elements, and a box length of \(L=1500 \unit{Mpc}\), with the observer positioned at the center.

The mock observations are generated according to the procedure described in section \ref{DRAWING_SIGNAL_SAMPLES}, with the underlying cosmological power-spectrum being calculated, with baryonic wiggles, following the prescription described in \citet{1998ApJ...496..605E} and \citet{1999ApJ...511....5E}. For these simulations  we assumed a standard \(\Lambda\)CDM cosmology with the set of cosmological parameters (\(\Omega_m=0.24\), \(\Omega_{\Lambda}=0.76\), \(\Omega_{b}=0.04\), \(h=0.73\), \(\sigma_8=0.74\), \(n_s=1\) ).

To generate the uncorrelated noise component, we assume a noise structure function \(n_i^{SF}=M_i\,F_i\,/\bar{\lambda}\), with \(\bar{\lambda}= 8.0\times10^{-3}\, L^3/N_{vox} \) in voxel space. Then we draw random Poission samples via the procedure described in section \ref{DRAWING_SIGNAL_SAMPLES}.

Note, that this is equivalent to drawing a galaxy distribution from the ensemble mean dark matter density, which has been obtained by averaging over all possible matter field realizations.

The survey properties are described by the galaxy selection function \(F_i\) and the observation Mask \(M_i\).
The selection function is given by:
\begin{equation}
\label{eq:MOCK_GAL_SELFUNC}
%F_i =\frac{r_i^b\,e^{-\left(\frac{r_i}{r_0}\right)^{\gamma}}}{((e^{\frac{ln(b/\gamma)}{\gamma}}\,r_0)^b\,e^{-exp(ln(b/\gamma)/\gamma)^{\gamma}})} \, ,
F_i = \left(\frac{r_i}{r_0}\right)^b\,\left(\frac{b}{\gamma}\right)^{-b/\gamma}\,e^{\frac{b}{\gamma}-\left(\frac{r_i}{r_0}\right)^{\gamma}} \, ,
\end{equation} 
where \(r_i\) is the comoving distance from the observer to the center of the \(i\)th voxel. For our simulation we chose parameters \(b=0.6\), \(r_0=500{\unit{Mpc}}\) and \(\gamma=2\).

In figure \nextext{\ref{fig:TEST_SEL_WIN}} we show the selection function together with the sky mask, which defines the observed regions in the sky. The two dimensional sky mask is given in sky coordinates of right ascension and declination.
The projection of this two dimensional mask into the three dimensional volume yields the three dimensional mask \(M_i\).

Two different projections of this generated mock galaxy survey are presented in figure \ref{fig:volume_rendering_observation} to give a visual impression of the artificial low resolution observation.

\begin{figure*}
\centering{\resizebox{1.\hsize}{!}{\includegraphics{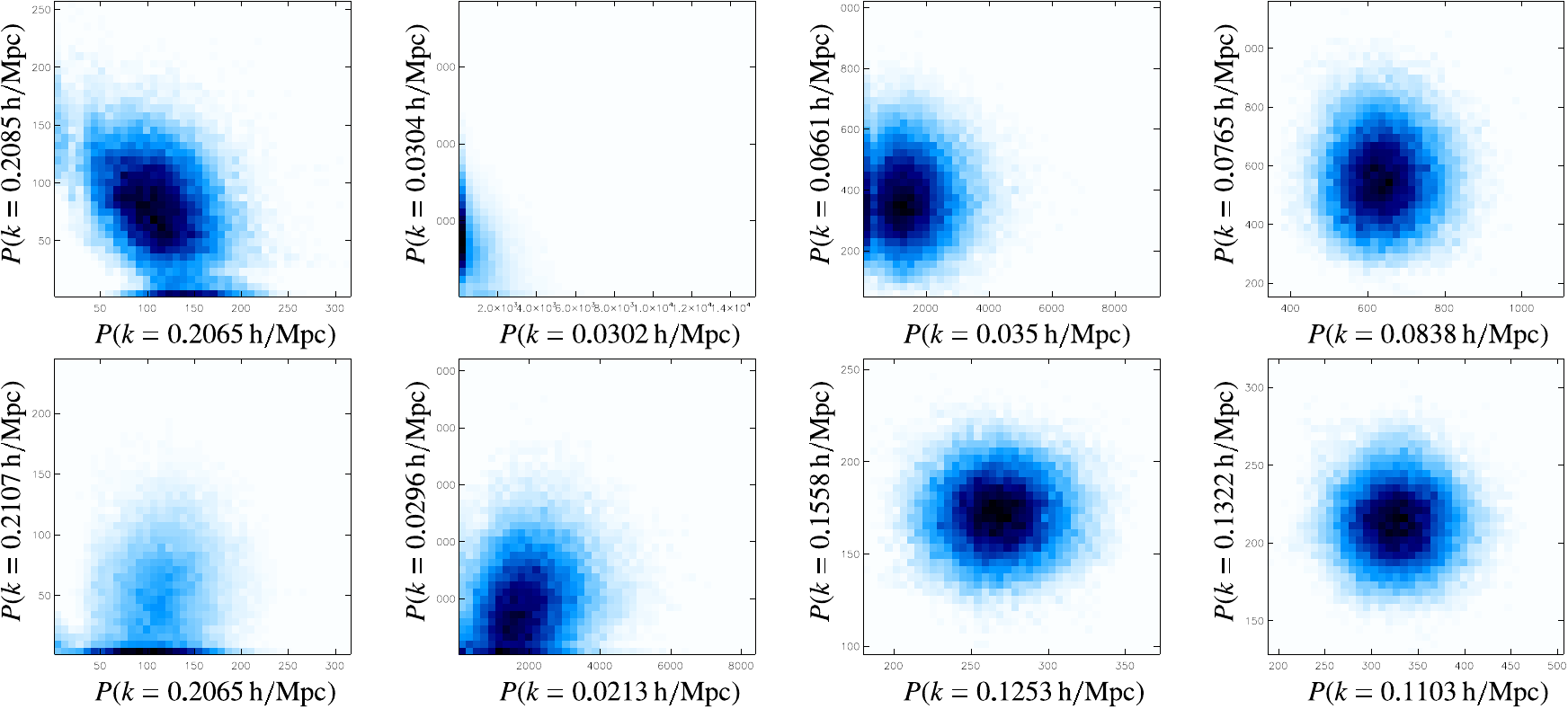}}}
\caption{2D marginalized posterior densities. Each plot shows the full joint posterior of the data, integrated over all dimensions except for the two shown.}
\label{fig:MARG_DENS_MI_PRIOR}
\end{figure*}

\subsection{Testing convergence and correlations}
The theory of Gibbs sampling states that the individual Gibbs samples converge to being samples from the joint probability distribution. However, the theory itself does not provide any criterion to detect when the samples start being samples from the joint probability distribution.
Therefore, this initial burn-in behavior has to be tested by experiments.
Further we will follow a similar analysis as described in \citet{2004ApJS..155..227E}, to estimate the convergence behavior and the correlation length of the Gibbs samples.

%Here we will analyze the convergence behavior, the correlation length and burn-in times, by following the standard test procedure as described in \citet{2004ApJS..155..227E}.

This analysis is important, as it allows us to understand the performance of our code in real world applications, which is particularly relevant for estimating how many Gibbs samples are required for an accurate power-spectrum estimation.

%Note, that the theory of Gibbs sampling states that the individual Gibbs samples converge to being samples from the joint probability distribution, but the theory itself does not provide any criterion to detect when the samples start being samples from the joint probability distribution.
%Therefore, this initial burn-in behavior has to be tested by experiments.

We study the initial burn-in behavior by setting up a simple experiment, in which we set the initial guess for the power-spectrum to be ten times larger in amplitude than the true underlying power-spectrum. Then \textsc{ARES} is applied to the mock observation, as described in the previous section \ref{SET_UP_MOCK_OBS}, to calculate a number of Gibbs iterations.
The obtained power-spectrum samples \(P_l^i\) of the \(i\)th iteration are then compared to the true power-spectrum via:
\begin{equation}
\label{eq:BURN_IN}
\xi_l^i = \frac{P_l^i-P_l^{true}}{P_l^{true}}\, ,
\end{equation}
where the \(P_l^{true}\) are the true power-spectrum coefficients, with which the mock dark matter signal was simulated.

The results for the \(\xi_l^i\) are shown in figure \ref{fig:BIT_CORRCOEFF}. Here, we observe a systematic drift of the successive power-spectra towards the true underlying power-spectrum. Further we see that the initial burn-in phase, for the kind of setup as demonstrated here, is rather short. The algorithm requires about \(30\) Gibbs iteration after which we can assume the samples to be samples from the joint probability distribution. Also note, that in this test, we do not observe any particular hysteresis for the poorly constrained large scale modes, meaning they do not remain at their initially set values. This demonstrates the ability of the code to handle correctly the mode coupling introduced by the sky cut.

However, it is clear that a poor initial guess invalidates a certain number of samples, especially at large scales, where the uncertainty due to the sky mask dominates.
For this reason, it is advantageous to initialize the Gibbs sampling chain with an initial guess, which is close to the true power-spectrum, to ensure short burn-in times.
As can be seen in figure \ref{fig:BIT_CORRCOEFF}, any bad initial guess would reveal itself by a systematic drift in the Gibbs chain, and can therefore be detected easily.

Next, we want to analyze the correlation of the individual Gibbs samples in the sequence.
%Another crucial point is the correlation of the individual Gibbs samples in the sequence.
This is a crucial point, as it permits us to estimate the number of independent samples, which can be obtained from a Gibbs chain of given length.

The correlation between sequential Gibbs samples can be best understood by reviewing the sampling algorithm.

A Gibbs sample of the joint probability distribution of signal and power-spectrum is obtained in two steps.
In the first step a Wiener reconstruction is performed, based on the assumption of a given power-spectrum, and the power lost due to noise filtering, masks and selection effects is replaced by a fluctuation term.
The signal obtained in this first sampling step mimics a full sky noise-less signal.
It is clear, that the power-spectrum of this signal is determined by the data in the high signal-to-noise region and by the assumed power-spectrum in the low signal-to-noise region.

In the second step the power-spectrum sample is generated, based on this full sky noise-less signal sample, obtained in the previous signal sampling step. The obtained power-spectrum then works as input power-spectrum to the next Gibbs step, and the iteration starts again.

In this fashion the Gibbs sampler performs a random walk in the multi-dimensional space of signal map and power-spectrum. 
The stepsize of the power-spectrum sampling step is solely determined by cosmic variance, and does not take into account the noise variance, as described in section \ref{Power_spectrum_sampling}.

However, we want to probe the full probability distribution, which includes both noise and cosmic variance.
This difference does not matter in the high signal to noise regime, since there cosmic variance will dominate the total variance, and for any practical purposes all Gibbs samples will be uncorrelated in this regime.
This, however, is not true in the low signal to noise regime. Since the random stepsize between two subsequent samples is determined only by the cosmic variance, and not by the much larger noise variance, two sequential samples will be strongly correlated. In this case a longer Gibbs sequence is required to produce uncorrelated samples.
%\begin{figure*}
%	\centering
%	{
%	\begin{picture}(100,240)

%	\put(-210,115){\rotatebox{90}{\(P(k)\, \left[(\rm{Mpc/h})^3\right]\)}}
%	\put(-90,0){\(k\) [h/Mpc]}
	
%	\put(55,115){\rotatebox{90}{\(P(k)\, \left[(\rm{Mpc/h})^3\right]\)}}
%	\put(185,0){\(k\) [h/Mpc]}

 % 	\put(-200,15){
%		\rotatebox{0}
%	{
%		\includegraphics[bb =103 78 433 397,width=0.45\textwidth,clip=true]{figs/MI_PRIOR/LOW_RES_SPEC_MIPRIOR.eps}
%	}
%	}

%	\put(65,15){
%			\rotatebox{0}
%	{
%			\includegraphics[bb =103 78 433 397,width=0.45\textwidth,clip=true]{figs/MI_PRIOR/HI_RES_SPEC_MIPRIOR.eps}
%	}
%	}
%\end{picture}
%}
%\caption{Power-spectrum estimates obtained from the Gibbs sampling chain for a low resolution (left panel) and a high resolution run (right panel). The black lines represent the ensemble mean of the sample set and the light gray and dark gray shaded regions denote the one and two sigma confidence regions respectively. Additionally we show the according input power-spectra. The blue line shows the cosmological power-spectrum from which the matter field realization was drawn, and the red line is the power-spectrum of this specific matter field realization.}
%	\label{fig:MI_PRIOR_SPECTRA}
%\end{figure*}
\begin{figure*}
\centering{\resizebox{1.\hsize}{!}{\includegraphics{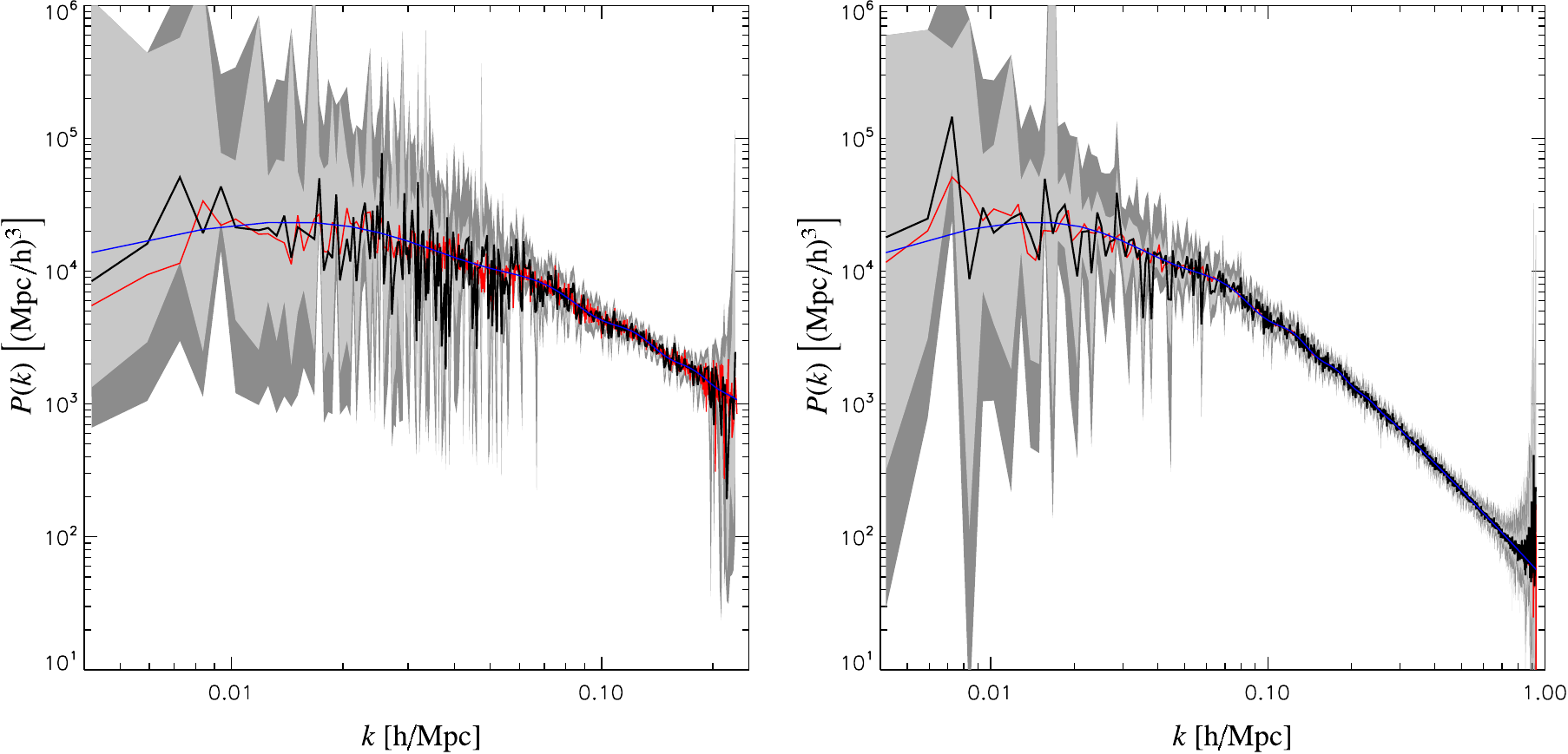}}}
\caption{Power-spectrum estimates obtained from the Gibbs sampling chain for a low resolution (left panel) and a high resolution run (right panel). The black lines represent the ensemble mean of the sample set and the light gray and dark gray shaded regions denote the one and two sigma confidence regions respectively. Additionally we show the according input power-spectra. The blue line shows the cosmological power-spectrum from which the matter field realization was drawn, and the red line is the power-spectrum of this specific matter field realization.}
\label{fig:MI_PRIOR_SPECTRA}
\end{figure*}

Reducing the variance by introducing binning to the power-spectrum, as described in section \ref{hidden_prior}, will lead to even longer correlation length in the Gibbs chain. This simply means, the joint probability distribution will be sampled with a finer resolution in power-spectrum space.

We study this correlation effect by assuming the power-spectrum coefficients \(P_l\) of different modes \(l\) in the Gibbs chain to be independent and estimate their correlation in the chain by calculating the autocorrelation function:
\begin{equation}
\label{eq:CORR_COEFF}
C_l(n) =\left \langle  \frac{P^i_l-\left \langle P_l\right \rangle}{\sqrt{Var P_l}} \frac{P^{i+n}_l-\left \langle P_l\right \rangle}{\sqrt{Var P_l}} \right \rangle \, ,
\end{equation}
where \(n\) is the distance in the chain measured in iterations \citep[for a similar discussion in case of the CMB see][]{2004ApJS..155..227E}.

We can then define the correlation length of the Gibbs sampler as the distance in the chain \(n_C\) beyond which the correlation coefficient \(C_l(n)\) has dropped below \(0.1\). 

The results for the different modes \(l\) are presented in figure \ref{fig:BIT_CORRCOEFF}.
As one can see, the vast majority of the different Fourier modes have a correlation length of about \(n_C\sim 100\) Gibbs iterations. The rest of the modes show increasingly longer correlation length, which increases towards the highest frequencies contained in the box. For this reason, the Nyquist frequency has the longest correlation length.
%\begin{figure*}
%	\centering
%	{
%	\begin{picture}(100,240)

%	\put(-205,130){\rotatebox{90}{\(\xi_l^i\)}}
%	\put(-90,0){\(k\) [h/Mpc]}
	
%	\put(60,130){\rotatebox{90}{\(C_l(n)\)}}
%	\put(190,0){\(n\)}

 % 	\put(-200,15){
%		\rotatebox{0}
%	{
%		\includegraphics[bb =104 78 433 395,width=0.45\textwidth,clip=true]{figs/INF_PRIOR/BIT_INF_PRIOR.eps}
%	}
%	}

%	\put(65,15){
%			\rotatebox{0}
%	{
%		\includegraphics[bb =95 78 431 395,width=0.459\textwidth,clip=true]{figs/INF_PRIOR/CORR_COEFF_INF_PRIOR.eps}
%	}
%	}
%\end{picture}
%}
%\caption{Same as figure \ref{fig:BIT_CORRCOEFF} but for low resolution runs with the inverse gamma prior.}
%	\label{fig:BIT_CORRCOEFF_INF_PRIOR}
%\end{figure*}
\begin{figure*}
\centering{\resizebox{1.\hsize}{!}{\includegraphics{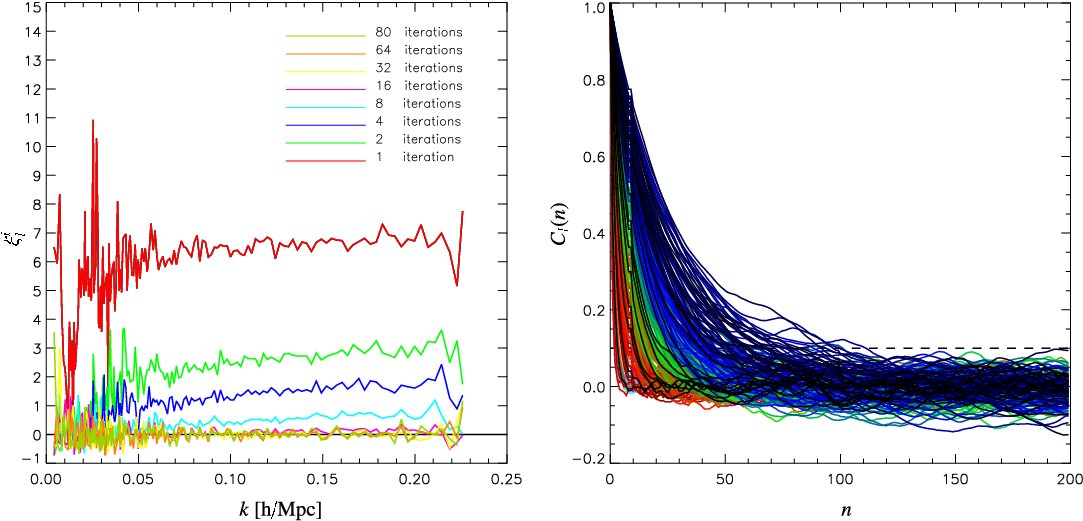}}}
\caption{Same as figure \ref{fig:BIT_CORRCOEFF} but for low resolution runs with the inverse gamma prior.}
\label{fig:BIT_CORRCOEFF_INF_PRIOR}
\end{figure*}
Especially for the highest frequencies towards the Nyquist frequencies, noise domination is only a partial explanation for the long correlation length. The far more important fact is, that the Nyquist frequency can in general not be resolved properly by the data, for example the Nyquist frequency is contained only once in the Fourier box. For this reason, the variance increases towards the Nyquist frequency. Note, that this effect arises from the technical implementation of the fast Fourier transform, which operates on a finite grid, and that we are in principle able to account for these technical effects of the analysis scheme itself. However, the long correlation length for the Nyquist frequencies will in general only provide a rare amount of independent estimates at the Nyquist frequency and therefore this must be taken into account in further analysis.

It is clear, that this effect shifts to higher frequencies as soon as the resolution of our analysis scheme is increased. This can be observed in the high resolution analysis discussed in the next section \ref{High_RES_SIM}.
Also note, that this effect can be cured by introducing an informative prior which will greately reduce the correlation length at these frequencies, as shown later in section \ref{Test_Informative_prior}. 

To study the marginalized posterior \(P_l\) distributions in more detail we plot their histograms in figure \ref{fig:MODE_STAT_CORR_MAT_MI_PRIOR}. It is worth mentioning that none of them is even approximately Gaussian.

Another crucial point to address is the question how well we were able to account for effects of the survey geometry. This information is contained in the correlation structure of the estimates.
Therefore, we can examine this effect by calculating the correlation matrix of the \(P_l\) estimates:
\begin{equation}
\label{eq:CORR_COEFF}
\mat{C}_{l\,l'} =\left \langle  \frac{P_l-\left \langle P_l\right \rangle}{\sqrt{Var P_l}} \frac{P_{l'}-\left \langle P_{l'}\right \rangle}{\sqrt{Var P_l}} \right \rangle \, ,
\end{equation}
where the ensemble averages are taken over \(40,000\) Gibbs samples.

We present the result in figure \ref{fig:MODE_STAT_CORR_MAT_MI_PRIOR}. It can be clearly seen that this correlation matrix has a very well defined diagonal structure, as expected from theory. The highest off-diagonal correlations have been measured to be less than \(20 \%\), and are found at the highest frequencies close to the Nyquist frequency. Figure \ref{fig:MODE_STAT_CORR_MAT_MI_PRIOR} also shows a blue band of anti-correlation around the diagonal. This anti-correlation indicates that the power-spectrum frequency resolution is higher than supported by the data. Since the data is fixed, and the mask couples neighbored Fourier modes, an increase in the power-spectrum amplitude \(P_l\) has to be compensated with a decrease in the neighboring power-spectrum amplitude \(P_{l+1}\) to have a good fit to the data. It is therefore possible, in a post-processing step, to reduce the frequency resolution of the estimated power-spectra until the anti-correlation vanishes. This is the idea behind the running median estimator, which will be presented in section \ref{OPERATIONS_GIBBS_SAMPLES}. 

However, since the posterior distributions for the \(P_l\) are non-Gaussian, the two point correlations do not contain all information. For this reason, we also demonstrate the marginalized posterior distribution for pairs of \(P_l\)s in figure \ref{fig:MARG_DENS_MI_PRIOR}, where we also show examples of maximally correlated and maximally anti-correlated modes.

Finally, we have plotted the full spectrum computed from our 40,000 sample run in figure \ref{fig:MI_PRIOR_SPECTRA}.
As can be seen, we chose a very high frequency resolution to reduce the correlation length of the Gibbs sampler in the low signal to noise regime. The estimated ensemble mean power-spectrum follows the true underlying power-spectrum, in particular the baryonic features. Towards the large scales, the uncertainty increases, as expected, since due to survey geometry and selection effects these scales are only poorly constrained by the data.

\subsection{High resolution Simulation}
\label{High_RES_SIM}
In the previous section, we performed a low resolution analysis to compute a sufficiently large set of samples to estimate the correlation behavior of our algorithm. However, such a large amount of samples is not necessarily required and computational time may be better invested in performing higher resolution analysis. Therefore, in this section we describe the results obtained from such a high resolution analysis.

The setup for this test is the same as described in section \ref{SET_UP_MOCK_OBS}, with the exception that here we use \(256^3\) voxels on an equidistant grid.

The main limitation for this test is CPU time, as a single sampling step takes about one hour.
For this reason, extremely long chains are not feasible. Hence, we run many, rather short, parallel Gibbs chains, as described in section \ref{Parallelization}. In this fashion we obtain an optimal speed up for this parallelization scheme.
The Gibbs sampler was run over 24 independently initialized chains and provided a total of \(4230\) samples.

We present the power-spectra obtained from the multiple-chain analysis in figure \ref{fig:MI_PRIOR_SPECTRA}. As can be seen there is overall good agreement with the realization specific power-spectrum and the ensemble mean estimate found by the Gibbs sampler. We also do not observe a detectable bias in any parts of the spectrum.

Towards the Nyquist frequency the uncertainty increases. This is expected, as towards the edges of the box, the number of modes decreases, and variance increases. Note, that in this fashion the method takes into account the uncertainty introduced by the analysis scheme itself, for example the Fourier space discretization of the FFT.

\begin{figure*}
\centering{\resizebox{1.\hsize}{!}{\includegraphics{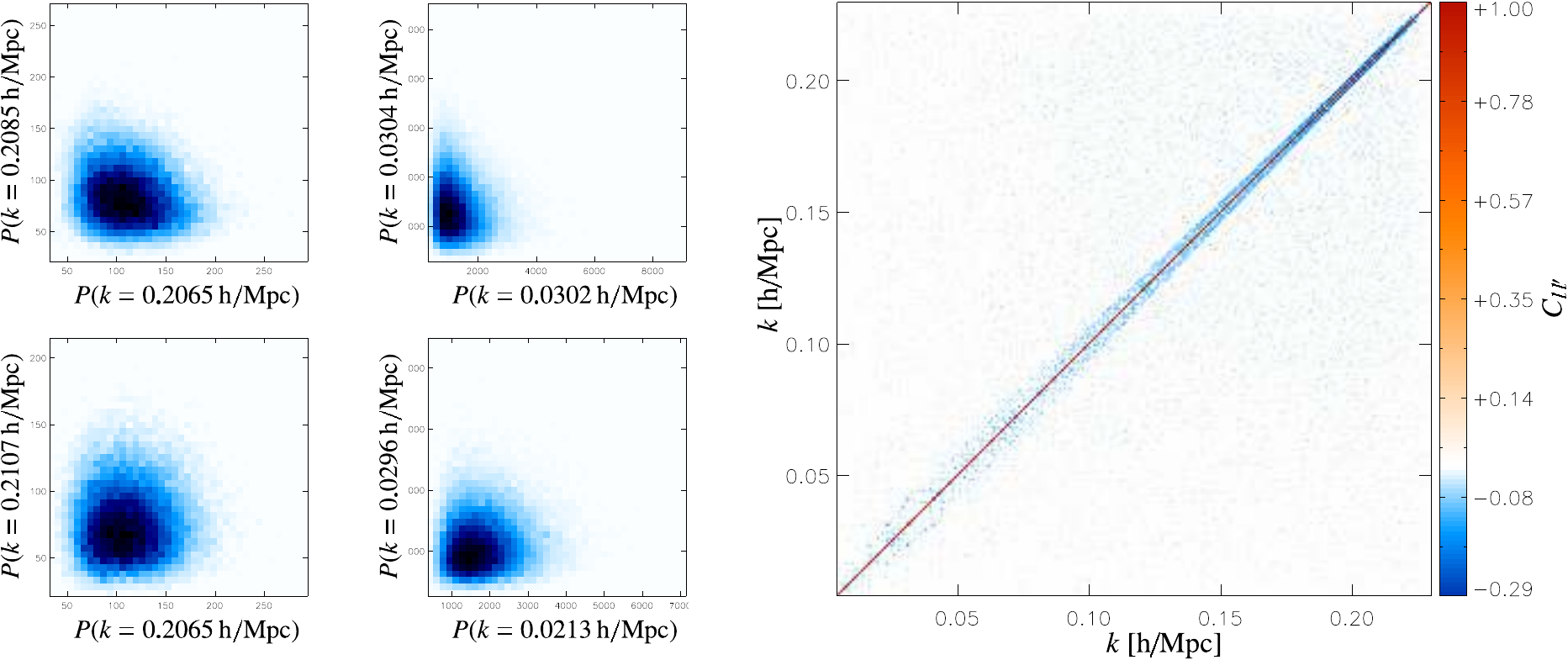}}}
\caption{The left panel shows four 2D marginalized posterior densities, which correspond to the maximally correlated and anti-correlated cases of the previous run. The right panel shows the power-spectrum covariance matrix estimated from the 40,000 Gibbs samples calculated with the informative prior.}
\label{fig:MARG_DENS_CORR_MAT_INF_PRIOR}
\end{figure*}

\subsection{Testing an informative Prior}
\label{Test_Informative_prior}
When trying to analyze real galaxy surveys, one is faced with the situation that due to sky cuts and selection effects, usually less than \(30\%\) of the volume is available for analysis. This affects mainly the estimate of the largest scales in the survey, as they are sparsely sampled, and therefore poorly constrained by the data.

In this situation the Jeffrey's prior, as a maximal ignorance prior, allows for large excursions from the expected true underlying power-spectrum. Since the Jeffrey's prior provides equal probability for all scales between zero and plus infinity, it also allows for power-spectrum values, which can be excluded on theoretical grounds, or by complementary experiments, which are more sensitive at the largest scales, such as CMB experiments.

From a Bayesian point of view, one could argue, that in the presence of a priori knowledge, a maximal ignorance prior is not the optimal choice. 

Rather than sampling the entire space of possible power-spectrum coefficients \(P_m\) with equal probability, it would be beneficial to preferably sample the region in which we expect the true power-spectrum to exist and allowing for larger excursions with smaller probability.

This would have the effect, that the region of interest would be sampled more densely, and therefore allowing for better power-spectrum estimates with the same amount of Gibbs samples. Also remember, that according to the discussion in section \ref{Blackwell-Rao_estimator}, the prior can be changed for any final post-processing analysis.

As an informative prior can lead to a more efficient sampling strategy in the presence of a priori knowledge, in the following we test \textsc{ARES} when employing the inverse gamma prior as described in section \ref{Informative_Prior}.

We base the inverse gamma prior on a flat power-spectrum guess, which was calculated according to \citet{1998ApJ...496..605E} and \citet{1999ApJ...511....5E}, without the baryonic wiggles. We explicitely do not incorporate the information on the baryonic features, to demonstrate that solely the data drives their estimate. Further, we choose \(n_m^{Prior}=5\), in order to make the prior sufficiently broad, while at the same time ensuring that it has finite variance.

With this prior choice we repeat the standard testing procedure as described in section \ref{SET_UP_MOCK_OBS}.

At first we test the initial burn-in time, by starting with a power-spectrum which is a factor \(10\) higher in amplitude than the underlying true power-spectrum.

The results for the according \(\xi_l^i\), described in equation (\ref{eq:BURN_IN}), are presented in figure \ref{fig:BIT_CORRCOEFF_INF_PRIOR}. It can be seen that the burn-in time is much shorter for the large scale modes, which are poorly constrained by the data. Also note, that the overall burn-in times for the power-spectra are comparable to those of the maximal ignorance prior case (see figure \ref{fig:BIT_CORRCOEFF}), indicating that these modes are not influenced by the informative inverse gamma prior.

The real advantage of the informative prior becomes obvious when analyzing the correlation length for the individual power-spectrum coefficients \(P_m\) in the Gibbs chain. Again we used a low resolution \(64^3\) Voxel simulation in order to estimate the correlation coefficients, as given by equation (\ref{eq:CORR_COEFF}). The results for this test are presented in figure \ref{fig:BIT_CORRCOEFF_INF_PRIOR}.

In comparison to figure \ref{fig:BIT_CORRCOEFF} it is clear, that the informative prior has a positive influence on the correlation length, which in this test are maximally on the order of hundred Gibbs iterations.

As discussed previously the long correlation length at the highest frequencies are mainly of technical nature, as the Nyquist frequencies cannot be properly represented in the finite Fourier box required for the FFT. This fact introduced artificial variance, which, however, our method can take into account. The informative prior helps in this situation, as it stabilizes these artificial fluctuations by prior information.

In addition, we observed a much better numerical behavior of our method when employing the informative prior, as the code does not run as frequently into numerically extreme regimes as with the maximum ignorance prior. This leads to a faster convergence of the conjugate gradient algorithm towards the desired accuracy.

Further, we also observe a better correction of survey geometry effects. The correlation function for the \(P_l\) is plotted in figure \ref{fig:MARG_DENS_CORR_MAT_INF_PRIOR}. The maximal correlation between different \(P_l\) in this test was less than \(10\%\), which is a clear improvement.

For comparison we also plot the 2D marginalized posterior densities, for the maximally correlated and anticorrelated modes in the maximum ignorance case.

\begin{figure*}
\centering{\resizebox{1.\hsize}{!}{\includegraphics{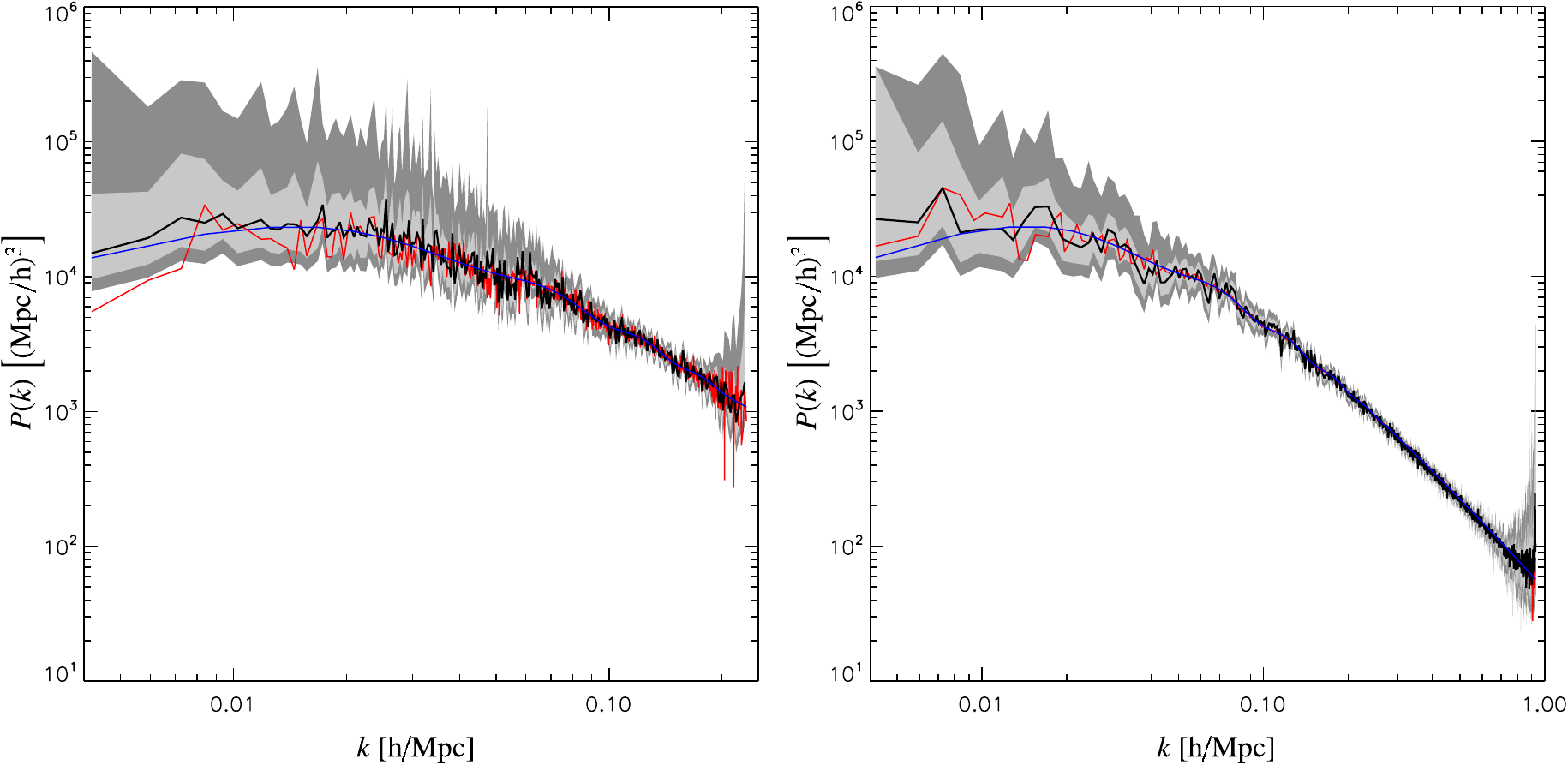}}}
\caption{Same as figure \ref{fig:mi_prior_SPECTRA} but for the inverse gamma informative prior.}
\label{fig:INF_PRIOR_SPECTRA}
\end{figure*}

Finally, we present the low and high resolution power-spectra for the informative prior in figure \ref{fig:INF_PRIOR_SPECTRA}. Note, that our prior did not contain any information on the baryonic oscillations. As can be clearly seen, the baryonic features have nicely been recovered, demonstrating, that our informative prior provided much less information than contained in the data.
%\begin{figure*}
%	\centering
%	{
%	\begin{picture}(100,240)

%	\put(-210,115){\rotatebox{90}{\(P(k)\, \left[(\rm{Mpc/h})^3\right]\)}}
%	\put(-90,0){\(k\) [h/Mpc]}
	
%	\put(55,115){\rotatebox{90}{\(P(k)\, \left[(\rm{Mpc/h})^3\right]\)}}
%	\put(185,0){\(k\) [h/Mpc]}

 % 	\put(-200,15){
%		\rotatebox{0}
%	{
%		\includegraphics[bb =103 78 424 397,width=0.4374\textwidth,clip=true]{figs/INF_PRIOR/LOW_RES_SPEC_INFPRIOR.eps}
%	}
%	}

%	\put(65,15){
%			\rotatebox{0}
%	{
%			\includegraphics[bb =103 78 433 397,width=0.45\textwidth,clip=true]{figs/INF_PRIOR/HI_RES_SPEC_INFPRIOR.eps}
%	}
%	}
%\end{picture}
%}
%\caption{Same as figure \ref{fig:MI_PRIOR_SPECTRA} but for the inverse gamma informative prior.}
%	\label{fig:INF_PRIOR_SPECTRA}
%\end{figure*}

\subsection{Testing with galaxy mock catalogs}
In this section, we describe the application of \textsc{ARES} to a mock galaxy survey based on the Millennium run \citep{2007MNRAS.375....2D}.

The intention of this exercise is two-fold. First we want to test \textsc{ARES} in a more realistic setup, where the intrinsic shot noise of the galaxy distribution is correlated with the underlying signal, which could not be tested with the Gaussian tests before. And second, we want to demonstrate that \textsc{ARES} is able to reconstruct the fully evolved non-linear matter distribution of the N-body simulation.

This mock galaxy survey consists of a set of comoving galaxy coordinates distributed in a \(500\) Mpc box. To obtain a realistic sky observation from this full sky galaxy sample, we virtually observe these galaxies through the sky mask and according to the selection function presented in figure \ref{fig:TEST_SEL_WIN}. The discrete galaxy distribution resulting from this mock observation is then sampled to a \(128^3\) equidistant grid.

To reduce gridding artifacts, such as aliasing power, we employ a supersampling technique as proposed in \citet{COSMO_DSP}. This allows us to accurately treat the mode coupling, and will yield a precision estimate of the power-spectra up to the highest frequencies contained in the box.

\begin{figure*}
\centering{\resizebox{1.\hsize}{!}{\includegraphics{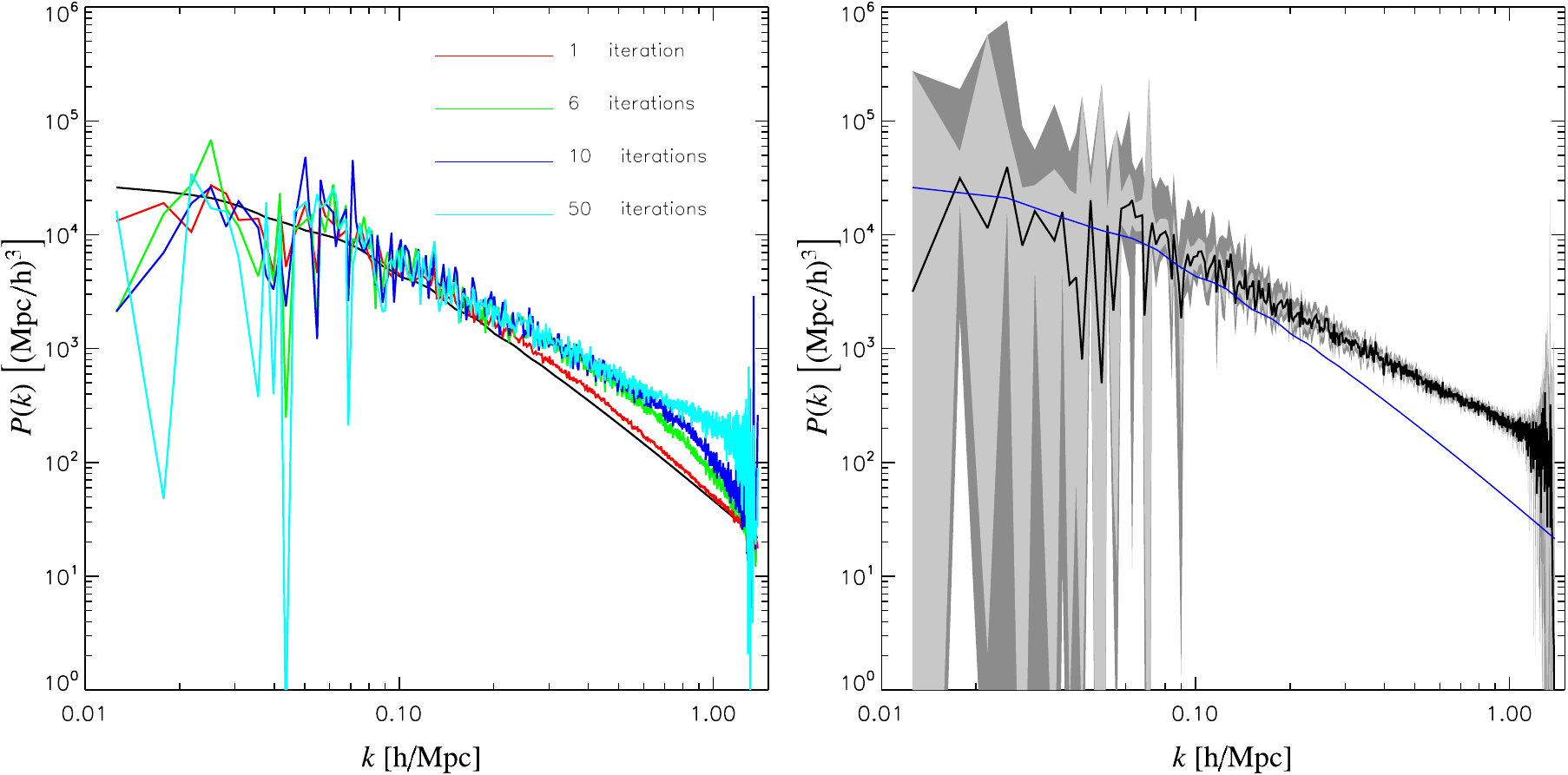}}}
\caption{The left panel shows successive power-spectrum samples during the burn-in period together with the initial guess (black line). The right panel shows the estimated power-spectrum of the Gibbs sampling analysis. The blue line denotes the initial power-spectrum guess, the black curve is the ensemble mean power-spectrum and the light gray and dark gray shaded regions represent the one sigma two sigma confidence regions respectively.}
	\label{fig:DELUCIA_SPEC}
\end{figure*}

Similarly to the method described in \ref{High_RES_SIM}, here we will run \(4\) independently initialized chains. Further, we employ the maximum ignorance Jeffrey's prior.
The galaxy distribution of this mock galaxy catalog follows the fully evolved non-linear matter distribution.
Nevertheless, we initialize the Gibbs sampler with the linear power-spectrum. Then the initial burn-in period, of about \(50\) samples,  is required to reach the non-linear power-spectrum. The systematic drift towards the correct power-spectrum is represented in figure \ref{fig:DELUCIA_SPEC}. This experiment nicely demonstrates that the Gibbs sampling approach is able to recover the non-linearities of the fully evolved matter density field.
At this point it is important to note, that the Wiener filter is a linear operation on the data, and as such leaves the statistics of the data intact. This has been demonstrated by \cite{KITAURA2009}, where they show, that the statistics of the reconstructed density field is consistent with a log-normal distribution, as expected for a non-linearly evolved matter distribution. This discussion clarifies, that the Wiener filter, or the Gibbs sampling approach as presented in this work, is very well able to capture the non-Gaussian characteristics of the density field.

However, in case one would like to perform a higher resolution analysis, it would be advantageous to initialize the Gibbs chain with a nonlinear power-spectrum guess, to yield even shorter burn-in times.

The ensemble averaged power-spectrum obtained from this run, together with the one and two sigma confidence regions, is also presented in figure
\ref{fig:DELUCIA_SPEC}. Here it can clearly be seen, that the recovered power-spectrum is consistent with the fully non-linearly evolved matter field. Towards the larger scales the uncertainty increase, which is due to the imposed survey geometry.

%As already mentioned earlier, the signal samples obtained by our Gibbs sampling procedure ore not of purely auxiliary character to create the power-spectrum samples, they are itself samples from the posterior probability distribution of the signal given the data. Therefore they provideinformation on the matter distribution which can be used for a variety of future applications.
%The advantage of the Gibbs sampling approach, as of any other Bayesian method, is the it does not only provide a single estimate, though it is certainly able to do so, but rather provides the entire probability distribution. We are therefore able to calculate any desired summary of the probability distribution such as the mean and variances.

\begin{figure*}
\centering{\resizebox{1.\hsize}{!}{\includegraphics{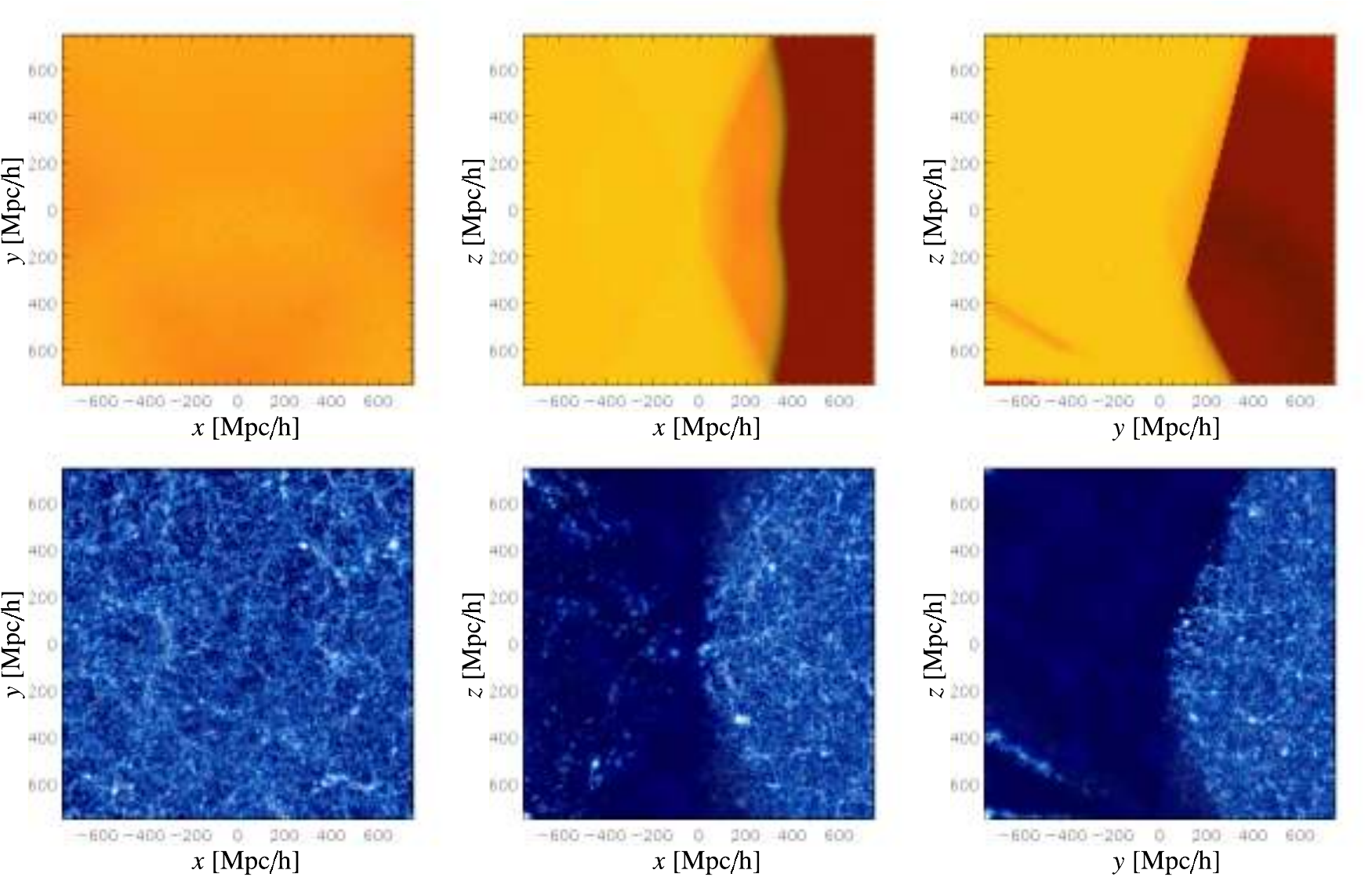}}}
\caption{Volume rendering of the ensemble variance (upper panels) and the ensemble mean (lower panels) obtained from the mock galaxy catalog analysis.}
\label{fig:DELUCIA_VARIANCE}
\end{figure*}

So far, in all test, we have focussed only on the recovery of the power-spectrum, and ignored the sample of reconstructed density fields.
Since the Gibbs sampler also provides samples from the matter density field posterior, we are able to calculate any desired statistical summary for the matter field reconstructions. The ability to provide uncertainty estimates for the recovered density fields will in general be valuable for further science based on the matter field estimates.
For this reason, in figure \ref{fig:DELUCIA_VARIANCE}, we present the estimated mean and variance maps obtained from the \(4000\) Gibbs samples.
As can be seen, the variance map clearly captures the features of the survey geometry and selection effects.% Additionally, one can see the influence of the shot noise contribution in high density regions.
With the set of Gibbs samples being available, all joint uncertainties can easily be propagated to the finally estimated quantities, such as the gravitational potential or large scale cosmic flows, by applying the according operation to the individual matter field samples.
The result of such a procedure then again yields a probability distribution in the final quantity, enabling us to provide uncertainty information for these quantities.

\section{Operations on the set of Gibbs samples}
\label{OPERATIONS_GIBBS_SAMPLES}
%In this section, we present two example applications operating on the set of Gibbs samples.
In this section, we present an example application operating on the set of Gibbs samples.
%We first demonstrate the possibility of seamlessly propagating the joint uncertainty to any final result in the case of the running median estimator. Then we perform a model comparison, employing the Blackwell-Rao estimator discussed in section \ref{Blackwell-Rao_estimator}.
%In the previous section we described the properties of the set of power-spectra obtained with \textsc{ARES}. As already described in the introduction, the outcome of a Bayesian analysis is not a single estimate but a probability distribution. In our specific case, this posterior probability distribution for the power-spectrum is represented by the power-spectrum Gibbs samples.

The outcome of our Bayesian method is not a single estimate but a Gibbs sample representation of the full posterior probability distribution for the power-spectrum coefficients. We are therefore able to propagate all uncertainties to any final result, simply by applying a post processing step to all Gibbs samples. The result of such an operation would again yield a probability distribution of the estimated final quantity.
%\begin{figure*}
%	\centering
%	{
%	\begin{picture}(100,240)

%	\put(-210,115){\rotatebox{90}{\(P(k)\, \left[(\rm{Mpc/h})^3\right]\)}}
%	\put(-90,0){\(k\) [h/Mpc]}
	
%	\put(55,115){\rotatebox{90}{\(P(k)\, \left[(\rm{Mpc/h})^3\right]\)}}
%	\put(190,0){\(k\) [h/Mpc]}

 % 	\put(-200,15){
%		\rotatebox{0}
%	{
%		\includegraphics[bb =103 78 424 397,width=0.45\textwidth,clip=true]{figs/DELUCIA/BIT_SPEC_DELUCIA.eps}
%	}
%	}

%	\put(65,15){
%			\rotatebox{0}
%	{
%			\includegraphics[bb =103 78 424 397,width=0.45\textwidth,clip=true]{figs/DELUCIA/SPEC_DELUCIA.eps}
%	}
%	}
%\end{picture}
%}
%\caption{The left panel shows successive power-spectrum samples during the burn-in period together with the initial guess (black line). The right panel shows the estimated power-spectrum of the Gibbs sampling analysis. The blue line denotes the initial power-spectrum guess, the black curve is the ensemble mean power-spectrum and the light gray and dark gray shaded regions represent the one sigma two sigma confidence regions respectively.}
%	\label{fig:DELUCIA_SPEC}
%\end{figure*}

As a simple demonstration, we apply a running median filter to the set of power-spectra, which will reduce the spectral resolution.

It is known, that the median is a better estimator of the typical value of a sample than the mean when there are large extraneous outliers in the sample \citep{STUART1994}. For this reason, we choose the median to estimate the mode power in a given frequency bin \(\Delta k_m\). Such a bin can be chosen to be large enough to smooth out any fluctuation below a certain scale. In our specific case we vary the bin width \(\Delta k_m\) with frequency, to allow for a logarithmic binning.

The median \(P^{\nu}_m\) of a set of power-spectrum amplitudes \(\{P_m\}\) contained within the frequency bin of width \(\Delta k_m\) around the mode \(k_m\) then satisfies the inequalities
\begin{equation}
\label{eq:MEDIAN_a}
\mathcal{P}\left(P_m\le P^{\nu}_m\right)\ge \frac{1}{2}
\end{equation}
and
\begin{equation}
\label{eq:MEDIAN_b}
\mathcal{P}\left(P_m\ge P^{\nu}_m\right)\ge \frac{1}{2} \, .
\end{equation}
The running median is then evaluated for every frequency in the power-spectrum sample.

We apply the running median to the set of power-spectrum samples obtained from the two Gaussian mock cases with the Jeffrey's and the inverse gamma prior. In doing so, we are able to calculate the mean and according uncertainty regions for the running median estimates. This effect has already been discussed in section \ref{hidden_prior}. Since the reduction of frequency resolution also decreases the amount of free parameters, the total variance decreases as well.

\begin{figure*}
\centering{\resizebox{1.\hsize}{!}{\includegraphics{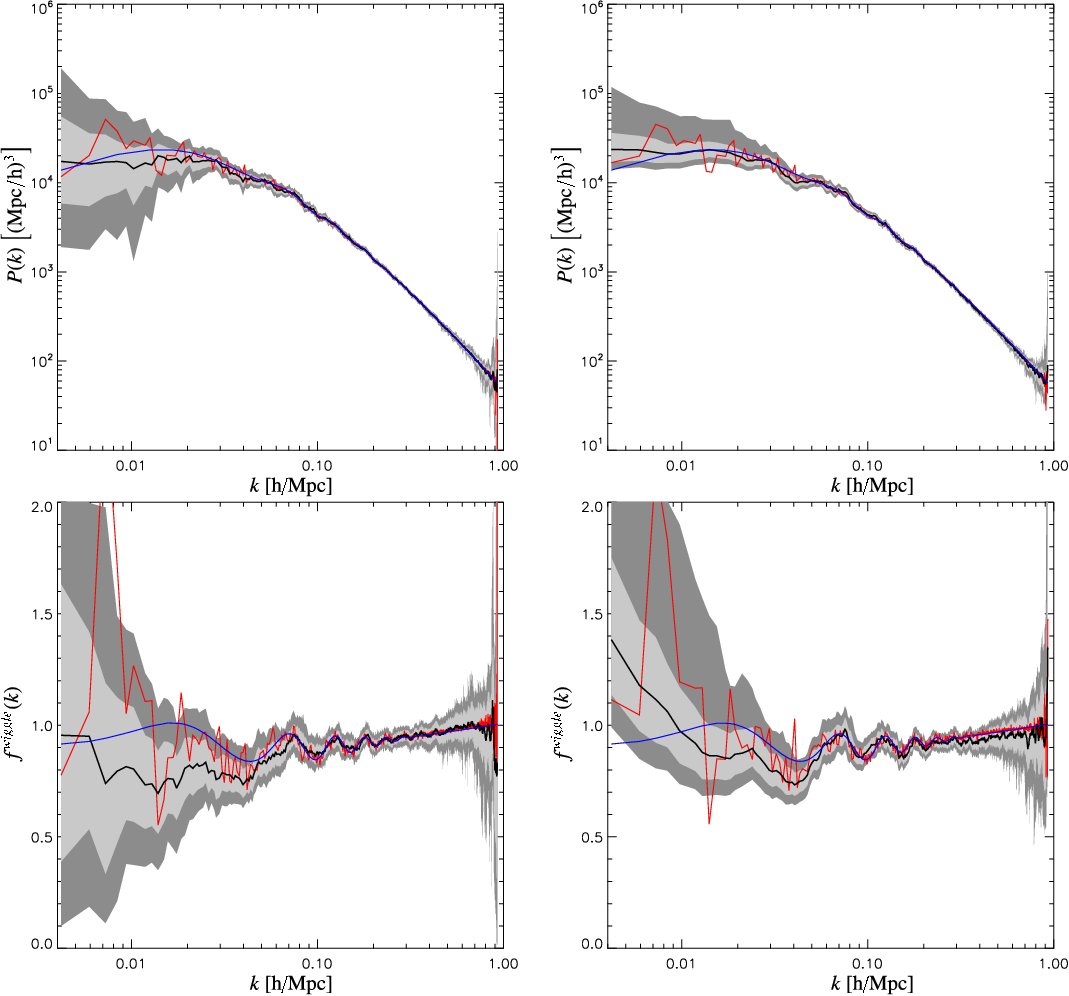}}}
\caption{Running median estimates of the power-spectra (upper panels) and the according wiggle functions (lower panels) for the set of Gibbs samples with the Jeffrey's prior (left panels), and the inverse gamma prior (right panels). The black lines represent the ensemble mean of the sample set and the light gray and dark gray shaded regions denote the one and two sigma confidence regions respectively. Additionally we show the according input power-spectra. The blue line shows the cosmological power-spectrum from which the matter field realization was drawn, and the red line is the power-spectrum of this specific matter field realization.}
\label{fig:RUNNING_MEDIAN}
\end{figure*}

The results of this experiment are demonstrated in figure \ref{fig:RUNNING_MEDIAN}. As one can easily see, the running median estimates are much smoother than the according Gibbs estimates. Also the reduction of frequency resolution by the running median estimator yields smaller confidence regions.

Finally we are interested in the recovery of the baryonic features in the power-spectrum. We therefore employ the common procedure of dividing the measured power-spectrum by one without baryonic wiggles \(P^{nowiggles}_m\). We then obtain the wiggle function as:
\begin{equation}
\label{eq:wiggle_function}
f^{wiggle}_m = \frac{P_m}{P^{nowiggles}_m} \, .
\end{equation}
Calculating the wiggle function for all Gibbs samples and applying the running median estimator to the set of wiggle functions will yield the distribution of wiggle functions. We then estimate the mean and the according one and two sigma confidence regions. The result of this calculation is presented in figure \ref{fig:RUNNING_MEDIAN} for the two Gaussian test cases. As expected, the variance towards the largest scales increases. Nevertheless, figure \ref{fig:RUNNING_MEDIAN} clearly demonstrates that the baryonic features have been recovered precisely by the Gibbs sampling approach.

This example nicely demonstrates that the uncertainty estimation can easily be transported to any final quantity estimated from the set of Gibbs samples. 

\section{Conclusion}
\label{Conclusion}
In this work, we presented \textsc{ARES}, a new Bayesian computer algorithm, designed to extract the full information on the two point statistics from any given probe of the three dimensional large scale structure.

The scientific aim of this algorithm is to provide an estimate of the power-spectrum posterior \(\mathcal{P}\left(\{P(k_i)\}|\{d_i\}\right)\), conditional on a data set, while accounting and correcting for all possible sources of uncertainties, such as survey geometry, selection effects and biases.
This is achieved by exploring the power-spectrum posterior \(\mathcal{P}\left(\{P(k_i)\}|\{d_i\}\right)\) via a Gibbs sampling approach.

While direct sampling from the power-spectrum posterior is not possible, it is possible to draw samples from the full joint posterior distribution \(\mathcal{P}\left(\{P(k_i)\},\{s_i\}|\{d_i\}\right)\) of the power-spectrum coefficients \(P(k_i)\) and the three dimensional matter density contrast amplitudes \(s_i\) conditional on a given set of data points \(\{d_i\}\).

The entire Gibbs sampling algorithm therefore consists of two basic sampling steps, in which first a full three dimensional Wiener reconstruction algorithm is applied to the data and then a power-spectrum is drawn from the inverse gamma distribution. 
In this fashion we obtain a set of power-spectrum and signal amplitude samples, which provide a full representation of the full joint posterior distribution \(\mathcal{P}\left(\{P(k_i)\},\{s_i\}|\{d_i\}\right)\).
The scientific output of this Bayesian method therefore is not a single estimate but a complete probability distribution, enabling us to calculate any desired statistical summary such as the mean, mode or variance.

We also demonstrated, that given a set of Gibbs samples, it is possible to provide an analytic approximation to the power-spectrum posterior \(\mathcal{P}\left(\{P(k_i)\}|\{d_i\}\right)\). This Blackwell-Rao estimator has an analytically appealing form enabling us to calculate any desired moment of the probability distribution in a simple analytic way.

In addition, since the full joint probability distribution is available, it is easy to carefully propagate all uncertainties through to the result of further post-processing analysis steps, such as parameter estimation. 

In this work, we focused on thoroughly testing the performance and behavior of our method by applying it to simulated data with controlled properties.
These tests were designed to highlight the problematic of survey geometry and selection effects, for the two cases of Gaussian random fields and a mock galaxy catalog based on the Millennium run.

One of the main goals of these tests was to build up intuition on the phenomenological behavior of the Gibbs sampling algorithm, estimating particularly issues, such as the correlation length of the Gibbs chain, burn-in and convergence times. The result of these tests is of special relevance, as it shows how long the Gibbs sampling chain has to run in order to produce a sufficient amount of independent samples. 

Through these experiments we found that the longest correlation lengths are dominated by the poorly constrained Nyquist modes of the box, which can be easily alleviated by imposing some prior knowledge on these modes. In doing so we found that the maximal correlation length for the Gibbs chain was on the order of hundred Gibbs samples. Thus, creating a large number of independent samples in a full scale data analysis is numerically very well feasible.

However, the most important result of these tests is, that our method is able to correct for artificial mode coupling due to the survey geometry and selection effects. This was tested by examining the correlation structure of the Gibbs samples, which showed that the maximal residual correlation can be reduced to less than \(10 \%\), demonstrating that this method correctly accounts for geometry effects.

The application of \textsc{ARES} to a galaxy mock catalog, based on the Millennium run, demonstrated the ability of our method to capture the characteristics of the fully nonlinearly evolved matter field. This is owed to the fact, that the Wiener filter is a linear operation on the data, and as such does not destroy the intrinsic statistical characteristics of the data set.

Nevertheless, a full real data analysis of existing redshift surveys requires the treatment of additional systematic effects such as scale or luminosity dependent biases or redshift space distortion corrections, which we defer to future works.
However, the Bayesian framework, as presented here, can take all these effects naturally into account and treats them statistically fully consistent.
Beside the possibility to include various kinds of uncertainties, the Gibbs sampling approach also allows for a natural joint analysis of different data sets, taking into account the systematics of each individual data set.

In summary, we showed that \textsc{ARES} is a highly flexible and adaptive machinery for large scale structure analysis, which is able to account for a large variety of systematic effects and uncertainties. For this reason, \textsc{ARES} has the potential to contribute to the era of precision cosmology.

\section*{Acknowledgments}
We are grateful to Rainer Moll and Bj\"{o}rn Malte Sch\"{a}fer for usefull discussions and support with many valuable numerical gadgets. We also thank J\'{e}r\'{e}my Blaizot for several enlightening explanations and hints on the nature of galaxy distributions and the systematics of redshift surveys.
Further we thank the "Transregional Collaborative Research Centre TRR 33 - The Dark Universe" and the Marie Curie FP7 fellowship for the support of this work.

\bibliography{paper}
\bibliographystyle{mn2e}
%%% Local Variables: 
%%% mode: latex
%%% TeX-master: t
%%% End: 

\appendix
\section{Discrete Fourier transformation}
\label{Discrete_Fourier_transformation}
we define the numerical Fourier transformation as:

\begin{equation}
\label{eqn:FFT}
y_k=\hat{C}\sum_{j=0}^{N-1}x_je^{-2\pi j k \frac{\sqrt{-1}}{N}}
\end{equation}
and the inverse Fourier transform
\begin{equation}
\label{eqn:IFFT}
x_j=C\sum_{k=0}^{N-1}y_k e^{2\pi j k \frac{\sqrt{-1}}{N}} = C\sum_{k=-(\frac{N}{2}-1)}^{\frac{N}{2}}y_k e^{2\pi j k \frac{\sqrt{-1}}{N}}  \, .
\end{equation}
The normalization coefficients \(C\) and \(\hat{C}\) are chosen such that they fulfill the requirement:
\begin{equation}
\label{eqn:Normalisation_condition}
C\hat{C}=\frac{1}{N} \, .
\end{equation}

\section{CHANGE TO FFT REPRESENTATION}
\label{CHANGE_TO_FFT_REPRESENTATION}
However, for computational convenience, we would rather like to consider the set of coefficients \(\hat{\hat{\mat{S}}}_{kl}\), which are the matrix coefficients of the Fourier transformed signal covariance matrix \(\mat{S}\). The signal covariance matrix coefficients \(\mat{S}_{ij}\) are related to the Fourier matrix coefficients \(\hat{\hat{\mat{S}}}_{kl}\) via the discrete Fourier transform defined in Appendix \ref{Discrete_Fourier_transformation}:
\begin{eqnarray}
\label{eq:FT_SIGNAL_COV}
\mat{S}_{ij} &=& C^2\,\sum^{N-1}_{k=0} \sum^{N-1}_{l=0}  e^{2\pi i k \frac{\sqrt{-1}}{N}}\, \hat{\hat{\mat{S}}}_{kl}\, e^{-2\pi j l \frac{\sqrt{-1}}{N}}  \nonumber \\
&=& C^2\,\sum^{N-1}_{k=0} \sum^{N-1}_{l=0}  e^{2\pi \frac{\sqrt{-1}}{N}(i\,k - j\,l )}\, \hat{\hat{\mat{S}}}_{kl}  \, ,
\end{eqnarray}
where \(N\) is the number of Volume cells which we consider in our calculation, and \(C\) is the normalization of the discrete Fourier Transform. The matrix elements of the Jacobi matrix for this coordinate transformation can then be written as:
\begin{eqnarray}
\label{eq:FT_SIGNAL_COV}
\mat{\mathcal{J}}_{(ij)\,(mn)}=\frac{\partial \mat{S}_{ij}}{\partial \hat{\hat{\mat{S}}}_{mn}} &=& C^2\,\sum^{N-1}_{k=0} \sum^{N-1}_{l=0}  e^{2\pi \frac{\sqrt{-1}}{N}(i\,k - j\,l )}\, \frac{\partial \hat{\hat{\mat{S}}}_{kl}}{\partial \hat{\hat{\mat{S}}}_{mn}}  \nonumber \\
&=& C^2\,\sum^{N-1}_{k=0} \sum^{N-1}_{l=0}  e^{2\pi \frac{\sqrt{-1}}{N}(i\,k - j\,l )}\, \delta^K_{km} \delta^K_{ln} \nonumber \\
&=& C^2\, e^{2\pi \frac{\sqrt{-1}}{N}(i\,m - j\,n )}\nonumber \\
&=& C^2\, \mat{A}_{(ij)\,(mn)}\, ,
\end{eqnarray}
where we defined the transformation matrix \(\mat{A}_{(ij)\,(mn)}\equiv e^{2\pi \frac{\sqrt{-1}}{N}(i\,m - j\,n )}\). The norm of the Jacobi determinant can then be calculated in matrix notation as:
\begin{eqnarray}
\label{eq:JACOBI_DETERMINANT}
\left | \rm{det}\left( \mat{\mathcal{J}} \right) \right | &=& \left | \rm{det}\left( C^2\, \mat{A} \right) \right | \nonumber \\ 
&=& \left | \rm{det}\left( C^2\,\mathbb{I}_{N^4}\, \mat{A} \right) \right | \nonumber \\
&=& \left | \rm{det}\left( C^2\,\mathbb{I}_{N^4}\right) \right |\, \left |\rm{det}\left( \mat{A} \right)\right |   \nonumber \\
&=& C^{2N^4} \left |\,\rm{det}\left( \mat{A} \right)\right |   \, ,
\end{eqnarray}
where we made use of the fact, that the Jacobi matrix is a quadratic \( N^2 \times N^2\) matrix, which is true for usual Fast Fourier Transform implementations.
Since \(\mat{A}\) is a unitary matrix:
\begin{eqnarray}
\label{eq:UNITARY_A}
\left ( \mat{A}_{(ij)\,(mn)} \right )^{\dagger} &=&  \mat{A}^*_{(mn)\,(ij)}  \nonumber \\ 
&=& e^{-2\pi \frac{\sqrt{-1}}{N}(m\,i - n\,j )} \nonumber \\
&=& e^{-2\pi \frac{\sqrt{-1}}{N}(i\,m - j\,n )}   \nonumber \\
&=& \mat{A}^{-1}_{(ij)\,(mn)}   \, ,
\end{eqnarray}
the norm of the determinant \(\rm{det}\left( \mat{A} \right)\) is unity. Therefore, the norm of the Jacobi determinant is just a constant:
\begin{equation}
\label{eq:JACOBI_DETERMINANT_final}
\left | \rm{det}\left( \mat{\mathcal{J}} \right) \right | = C^{2N^4} \, .
\end{equation}

With this definition we can perform the change of coordinates and express the probability distribution given in equation (\ref{eq:COND_INDEP_PS_set}) in terms of the set of the matrix coefficients \(\hat{\hat{\mat{S}}}_{kl}\):
\begin{equation}
\label{eq:COND_INDEP_PS_FT}
{\cal P}(\{\hat{\hat{\mat{S}}}_{kl}\}|\{s_i\}) = C^{2N^4} \, {\cal P}(\{\mat{S}_{ij}\}|\{s_i\}) = C^{2N^4} \frac{{\cal P}(\{\hat{\hat{\mat{S}}}_{kl}\})}{{\cal P}(\{s_i\})} {\cal P}(\{s_i\}|\{\hat{\hat{\mat{S}}}_{kl}\}) \, ,
\end{equation}
where \( {\cal P}(\{\hat{\hat{\mat{S}}}_{kl}\}) \) is the signal covariance matrix prior in Fourier space, and \({\cal P}(\{s_i\}|\{\hat{\hat{\mat{S}}}_{kl}\})\) is given as:
\begin{eqnarray}
\label{eq:FT_likelihood_PS}
{\cal P}(\{s_i\}|\{\hat{\hat{\mat{S}}}_{kl}\}) &=&  {\mathcal P}(\{s_i\}|\mat{S})\nonumber \\ 
&=& \frac{1}{\sqrt{\rm{det}\left(2\pi \mat{S}\right)}}e^{-\frac{1}{2}\sum_i\sum_j s_i \mat{S_{ij}}^{-1}s_j}  \nonumber \\
&=& \frac{1}{\sqrt{\rm{det}\left(2\pi \mat{S}\right)}}e^{-\frac{1}{2}\sum_i\sum_j s_i C^2\,\sum^{N-1}_{k=0} \sum^{N-1}_{l=0}  e^{2\pi \frac{\sqrt{-1}}{N}(i\,k - j\,l )}\, \hat{\hat{\mat{S}}}^{-1}_{kl}s_j}    \nonumber \\
&=& \frac{1}{\sqrt{\rm{det}\left(2\pi \mat{S}\right)}}e^{-\frac{C^2}{2}\sum^{N-1}_{k=0} \sum^{N-1}_{l=0} \sum_i s_i \,  e^{2\pi \frac{\sqrt{-1}}{N}\,i\,k}\, \hat{\hat{\mat{S}}}^{-1}_{kl} \sum_j \,s_j\,e^{-2\pi \frac{\sqrt{-1}}{N}\,j\,l }}   \nonumber \\
&=& \frac{1}{\sqrt{\rm{det}\left(2\pi \mat{S}\right)}}e^{-\frac{C^2}{2\,\hat{C}^2}\sum^{N-1}_{k=0} \sum^{N-1}_{l=0} \hat{s}^*_k \,  \hat{\hat{\mat{S}}}^{-1}_{kl} \,\hat{s}_l\,}   \, ,
\end{eqnarray}
with \(\hat{s}_l\) being the signal coefficients of the discrete Fourier transformed signal.
To determine the determinant \(\rm{det}\left(2\pi \mat{S}\right)\) in Fourier space, we rewrite equation (\ref{eq:FT_SIGNAL_COV}) as a matrix product:
\begin{eqnarray}
\label{eq:FT_SIGNAL_COV_mat}
\mat{S}_{ij} &=& \sum^{N-1}_{k=0} \sum^{N-1}_{l=0}  C\,e^{2\pi i k \frac{\sqrt{-1}}{N}}\, \hat{\hat{\mat{S}}}_{kl}\, C\,e^{-2\pi j l \frac{\sqrt{-1}}{N}}  \nonumber \\
&=& \sum^{N-1}_{k=0} \sum^{N-1}_{l=0}  C A_{ik}\, \hat{\hat{\mat{S}}}_{kl}\, C\,A^*_{lj}  \,
\end{eqnarray}
where we defined the unitary matrix \( A_{ik}\equiv\,e^{2\pi i k \frac{\sqrt{-1}}{N}}\). The determinant can then be evaluated in matrix notation as:
\begin{eqnarray}
\label{eq:FT_DETERMINANT}
\rm{det}\left(2\pi \mat{S}\right) &=& \rm{det}\left(2\pi C^2 A \hat{\hat{\mat{S}}} A^{-1} \right) \nonumber \\ 
&=& \rm{det}\left(C^2\,\mathbb{I}_{N^2}  A  2\pi \hat{\hat{\mat{S}}} A^{-1} \right) \nonumber \\ 
&=& \rm{det}\left(C^2\,\mathbb{I}_{N^2} \right)  \rm{det}\left(A\right)  \rm{det}\left(2\pi \hat{\hat{\mat{S}}}\right) \rm{det}\left(A^{-1} \right) \nonumber \\
&=& C^{2\,N^2}\rm{det}\left(2\pi \hat{\hat{\mat{S}}}\right) \, .
\end{eqnarray}

\section{Wiener Variance}
\label{WIENER_VARIANCE}
In section \ref{DRAWING_SIGNAL_SAMPLES} we described, that a signal sample can be obtained by adding a fluctuation term \(y_i\) to the Wiener mean reconstruction \(m_i\). According to equations (\ref{eq:FLUCTUATION_TERM}) and (\ref{eq:MOCK_DATA})  we can rewrite the fluctuation term as:
\begin{eqnarray}
\label{eq:RE_FLUCTUATION_TERM}
y_i&=& s_i^*-\sum_j \mat{D}^{-1}_{ij}\sum_m \sum_l \mat{K}_{mj} \mat{N}^{-1}_{ml} d_l^* \nonumber \\ 
&=& s_i^*-\sum_j \mat{D}^{-1}_{ij}\sum_m \sum_l \mat{K}_{mj} \mat{N}^{-1}_{ml} \left(\sum_k \, K_{lk}\, s_k^* + \epsilon_l^*\right) \nonumber \\
&=& \sum_j \mat{D}^{-1}_{ij}\left[\sum_k \, D_{jk}\, s_k^*-\sum_m \sum_l \mat{K}_{mj} \mat{N}^{-1}_{ml} \left(\sum_k \, K_{lk}\, s_k^* + \epsilon_l^*\right) \right] \nonumber \\
&=& \sum_j \mat{D}^{-1}_{ij}\left[\sum_k \,\left( D_{jk} -\sum_m \sum_l \mat{K}_{mj} \mat{N}^{-1}_{ml} \, K_{lk}\, \right)\, s_k^*-\sum_m \sum_l \mat{K}_{mj} \mat{N}^{-1}_{ml}  \epsilon_l^* \right] \nonumber \\
&=& \sum_j \mat{D}^{-1}_{ij}\left[\sum_k \, S^{-1}_{jk}\, s_k^*-\sum_m \sum_l \mat{K}_{mj} \mat{N}^{-1}_{ml}  \epsilon_l^* \right] \, .
\end{eqnarray}
From this we immediately see that the mean of \(y_i\):
\begin{equation}
\label{eq:RE_FLUCTUATION_TERM_MEAN}
\langle y_i \rangle =  \sum_j \mat{D}^{-1}_{ij}\left[\sum_k \, S^{-1}_{jk}\, \langle s_k^*\rangle-\sum_m \sum_l \mat{K}_{mj} \mat{N}^{-1}_{ml}  \langle\epsilon_l^*\rangle \right]=0 \, 
\end{equation}
vanishes by construction, as \(\langle s_k^*\rangle=\langle\epsilon_l^*\rangle=0\).  Note, that the signal and noise covariance are given by:
\begin{equation}
\label{eq:RE_FLUCTUATION_TERM_MEAN}
\langle s_i^* s_j^* \rangle =  \mat{S}_{ij} \, 
\end{equation}
and
\begin{equation}
\label{eq:RE_FLUCTUATION_TERM_MEAN}
\langle \epsilon_i^* \epsilon_j^* \rangle =  \mat{N}_{ij} \, . 
\end{equation}
Also note that , by construction, there is no correlation between the two random variates \(s_i^*\) and \(\epsilon_i^*\)  as they have been created by two independent random processes. Therefore, we yield:
\begin{equation}
\label{eq:RE_FLUCTUATION_TERM_MEAN}
\langle s_i^* \epsilon_j^* \rangle =  0 \,. 
\end{equation}
Then we can calculate the variance as:
\begin{eqnarray}
\label{eq:RE_FLUCTUATION_TERM}
\langle y_i y_q \rangle&=& \left \langle \sum_j \mat{D}^{-1}_{ij}\left[\sum_k \, S^{-1}_{jk}\, s_k^*-\sum_m \sum_l \mat{K}_{mj} \mat{N}^{-1}_{ml}  \epsilon_l^* \right]  \, \sum_r \mat{D}^{-1}_{qr}\left[\sum_s \, S^{-1}_{rs}\, s_s^*-\sum_t \sum_u \mat{K}_{tr} \mat{N}^{-1}_{tu}  \epsilon_u^* \right] \right \rangle \nonumber \\
&=& \sum_{jr} \mat{D}^{-1}_{ij}  \mat{D}^{-1}_{qr} \left[\sum_{ks} \, S^{-1}_{jk} S^{-1}_{rs}\, \langle s_k^*\,s_s^* \rangle +\sum_{mltu} \mat{K}_{mj} \mat{N}^{-1}_{ml} \mat{K}_{tr} \mat{N}^{-1}_{tu}  \langle  \epsilon_l^* \epsilon_u^* \rangle \right]   \nonumber \\
&=& \sum_{jr} \mat{D}^{-1}_{ij}  \mat{D}^{-1}_{qr} \left[\sum_{ks} \, S^{-1}_{jk} S^{-1}_{rs}\, \mat{S}_{ks} +\sum_{mltu} \mat{K}_{mj} \mat{N}^{-1}_{ml} \mat{K}_{tr} \mat{N}^{-1}_{tu}  \mat{N}_{lu} \right]   \nonumber \\
&=& \sum_{jr} \mat{D}^{-1}_{ij}  \mat{D}^{-1}_{qr} \left[\sum_{k} \, S^{-1}_{jk} \delta^K_{rk} +\sum_{mlt} \mat{K}_{mj} \mat{N}^{-1}_{ml} \mat{K}_{tr} \delta^K_{tl} \right]   \nonumber \\
&=& \sum_{jr} \mat{D}^{-1}_{ij}  \mat{D}^{-1}_{qr} \left[S^{-1}_{jr} +\sum_{ml} \mat{K}_{mj} \mat{N}^{-1}_{ml} \mat{K}_{lr}  \right]   \nonumber \\
&=& \sum_{jr} \mat{D}^{-1}_{ij}  \mat{D}^{-1}_{qr} \mat{D}_{jr}  \nonumber \\
&=& \sum_{r}   \mat{D}^{-1}_{qr}  \delta^K_{ir}  \nonumber \\
&=& \mat{D}^{-1}_{iq} \, .
\end{eqnarray}
Therefore, the additive fluctuation term \(y_i\) provides the correct variance.

\bsp

\label{lastpage}

\end{document}